\documentclass[%
 aip,
 amsmath,amssymb,
reprint,%
]{revtex4-1}

\usepackage{graphicx}
\usepackage{dcolumn}
\usepackage{bm}

\usepackage{framed,multirow}
\usepackage{amssymb}
\usepackage{amsmath}
\usepackage{mathtools}
\usepackage{latexsym}
\usepackage{eucal}
\usepackage{siunitx}
\usepackage{url}
\usepackage{xcolor}

\usepackage[utf8]{inputenc}
\usepackage[T1]{fontenc}
\usepackage{mathptmx}
\usepackage{etoolbox}
\usepackage[english]{babel}
\makeatletter
\def\@email#1#2{%
 \endgroup
 \patchcmd{\titleblock@produce}
  {\frontmatter@RRAPformat}
  {\frontmatter@RRAPformat{\produce@RRAP{*#1\href{mailto:#2}{#2}}}\frontmatter@RRAPformat}
  {}{}
}%
\makeatother

\newcommand{\tbf}[1]{\textbf{#1}}
\newcommand{\mrm}[1]{\mathrm{#1}}
\newcommand{\mbf}[1]{\mathbf{#1}}

\newcommand{\abs}[1]{\|{#1}\|}

\newcommand{\unitvec}[1]{\hat{\mathbf{#1}}}

\DeclareMathOperator\erfc{erfc}

\newcommand\bovermat[2]{%
  \makebox[0pt][l]{$\smash{\overbrace{\phantom{%
    \begin{matrix}#2\end{matrix}}}^{\text{#1}}}$}#2}

\newcommand{\kb}{ k_{\mathrm{B}} }
\renewcommand{\thefigure}{\arabic{figure}}
\begin{document}

\preprint{ }

\title[ ]{ Immersed boundary method for dynamic simulation of polarizable colloids of arbitrary shape in explicit ion electrolytes}
\author{Emily Krucker-Velasquez }
\author{James W. Swan}
\affiliation{%
 Department of Chemical Engineering, Massachusetts Institute of Technology, Cambridge, Massachusetts 02139, USA
}%
\author{Zachary Sherman }
\email{zsherm@uw.edu}
\affiliation{ Department of Chemical Engineering, University of Washington, Seattle, Washington 98195, USA}
\date{\today}
\begin{abstract}

We develop a computational method for modeling electrostatic interactions of arbitrarily-shaped, polarizable objects on colloidal length scales, including colloids/nanoparticles, polymers, and surfactants, dispersed in explicit ion electrolytes and nonionic solvents. Our method computes the nonuniform polarization charge distribution induced in a colloidal particle by both externally applied electric fields and local electric fields arising from other charged objects in the dispersion.  This leads to expressions for electrostatic energies, forces, and torques that enable efficient molecular dynamics and Brownian dynamics simulations of colloidal dispersions in electrolytes, which can be harnessed to accurately predict structural and transport properties. We describe an implementation in which colloidal particles are modeled as rigid composites of small spherical beads that tessellate the surface of the particle. The electrostatics calculations are accelerated using a spectrally-accurate particle-mesh-Ewald technique implemented on a graphics processing unit (GPU) and regularized such that the electrostatic calculations are well-defined even for overlapping bodies. We demonstrate the effectiveness of this approach through a series of calculations, including the induced dipole moments and forces for one, two, and lattices of spherical colloids in an electric field; the induced dipole moment and torque for anisotropic particles in an electric field; the equilibrium distribution of ions in the double layer around charged colloids; the dynamics of charged colloids; and ions in the double layer around a polarizable colloid exposed to an electric field. 
\end{abstract}

\maketitle


\section{Introduction}
\indent The equilibrium and dynamic properties of colloidal suspensions are critical for a wide variety of industrial applications. Colloidal suspensions can be as simple as gold particles in water or be present in more sophisticated structures such as micelles, vesicles, nanocapsules, and dendritic polymers to name a few \cite{caruso2003}(herein simply colloidal particles). When immersed in a dielectric fluid, colloidal particles often acquire charges. The resulting many-body electrostatic interactions inform the macroscopic properties of the suspension.\cite{israelachvili2011} For example, charged colloids attract oppositely charged species to the colloid-fluid interface, giving rise to what is commonly referred to as the ``electric double layer'' (EDL).\cite{Russel1989} The thickness of this microscopic electric double layer greatly influences the strength of interactions among the colloids and consequently the thermodynamic stability of the suspension.\cite{Russel1989,israelachvili2011}  Furthermore, charged colloids and their respective ionic clouds behave as quasiparticles in the presence of phoretic forces; thus, the thickness of the EDL controls the electrophoretic mobility of the particles.\cite{kjellander_dressed_1997,kjellander_2001,Russel1989} 
Alternatively, polarization effects can lead to emergent transport processes. This is the case for polarizable colloidal particles in an external electric field. The field induces a nonuniform surface charge distribution giving rise to the nonlinear electrokinetic phenomenon known as "induced charge electro-osmosis". \cite{Bazant2004}\\
\indent Several continuum theories have been developed to describe these effects, the most common of which are Debye-Huckel theories for equilibrium properties and the ``standard model'' of electrokinetics for dynamic properties,\cite{Russel1989} both of which consider noninteracting point ions. Because these theories ignore ion-ion correlations and excluded volume interactions, they are inaccurate for a range of conditions important in colloidal suspensions, including strongly charged colloids,\cite{israelachvili2011} concentrated electrolytes, and ionic liquids.\cite{Gebbie2017} \\
\indent Particle-based simulations can be tailored to include such ion-ion correlations and excluded volume interactions. For example, several molecular dynamics, multiparticle collision dynamics, and Monte Carlo techniques have been developed to model electrostatic interactions in colloidal systems \cite{Ulberg1993,delGado2014, chen_nanoparticle_2021,Eshraghi2018,Sidhu2018,Padidela2017}. Nevertheless, the majority of previous computational work considers only pairwise additive Coulombic potentials between charged species and does not consider the importance of three-body and higher-order (many-body) interactions. While useful for certain classes of problems, these simplifications disregard 1) charges that may be induced in polarizable bodies due to global or local electric fields and 2) many-body effects on the energies and forces of these polarizable bodies. These time- and configuration-dependent ``induced charge'' effects are important for accurately describing the structure of ionic double layers,\cite{outhwaite_influence_2011} stability and self-assembly of colloidal suspensions,\cite{Sherman2018,kang_double-layer_2008,zhang2015toward} and particle and fluid dynamics in electric fields.\cite{Bazant2004} \\
%
\indent In general, the level of ``coarse-graining'' strongly influences the level of accuracy of particle-based simulations. There are different levels of coarse-graining that can be used when simulating colloidal suspensions. The highest level of refinement at the length scale of our interest is present in conventional all-atom molecular dynamics. Here, each of the atoms in the suspension is replaced by appropriately scaled beads whose interparticle interactions are approximated via force fields. All-atom simulations, however, are prohibitively expensive when colloidal systems are studied because of the significant difference between the solvent and colloidal particles' sizes. A single ${10^{-8}\;\mrm{m}}$ colloid comprises approximately ${2\times10^{4}}$ particles.\cite{Winslow2019} Common solvents, such as water, are on the length scale of ${10^{-2}}$ the diameter of this colloidal particle. It would be necessary to simulate a number in the order of ${10^6}$ solvent molecules per unit volume of colloid. Thus, this level of detail hinders our ability to simulate even a small number of colloids. For example, if one considers simulating four colloidal particles in a tetrahedral packing density; this would be equivalent to solving ${10^{7}}$ equations of motion each time step and keeping track of the equivalent number of positions and velocities. \\ %
\indent In contrast, continuum models are highly coarse-grained. Continuum approximations, such as the Poisson-Boltzmann (PB) equation, commonly assume that colloidal particles lie in a solvent of constant permittivity and only experience the \emph{average} behavior of other neighboring charged species. In these models, polarization, many-body effects, and thermal fluctuations can noticeably increase the complexity of the formulation of the problem and preclude analytic progress. Therefore, these effects are often ignored in purely continuum models, and consequently, this simplified representation of charged soft matter systems is only accurate in a finite range of regimes. This range can broaden if thermal fluctuations are explicitly included in the simulation. \cite{Ladiges_Donev_2021}\\
\indent Within the level of detail present in all-atom simulations and continuum models, there are alternative schemes that allow us to reduce the computational cost of simulating at the colloidal scale. Brownian dynamics (BD), for example, can be used to describe the motion of discrete particles in a continuum Newtonian fluid. This significantly decreases the computational resources needed to track the positions and velocities of the solvent. When coupled with Stoke's equation (known as Stokesian Dynamics)\cite{Brady1988,elfring_brady_2022} this hybrid method can be used for a wide range of concentrations to numerically extract the dynamic and static properties of assembly particles whose radii range from $10^{-9}\; \mrm{m}$ (seen in the average hydrated diameter of sodium chloride in solution) to $10^{-3}\; \mrm{m}$,\cite{SwanGang2016, israelachvili2011} without the need to explicitly simulate the solvent molecules.\\
\indent Alternatively, different numerical methods can be used to solve Poisson's equation and compute the interparticle forces in electrostatically stabilized colloidal suspensions. Some of these numerical treatments are finite differences, finite elements, and finite boundary methods. However, this can be a difficult task due to the long-ranged and many-body nature of electrostatic interactions.\\ 
\indent These techniques to solve partial differential equations, however, require at least the discretization of the entire 2D space (finite difference and finite element methods discretize the entire 3D space), which is computationally costly. Various numerical schemes and approximations have emerged to overcome this problem. One of the simplest ways to avoid discretization of space is through effective interaction potentials. Particle-based simulations often treat the ions and colloids as discrete particles that interact through some effective Yukawa-type interaction potential (see, for example, ref. \cite{OveradeLaCruz2010} or ref. \cite{dobnikar2004} ). This circumvents the need to otherwise compute the electric potential in the simulation box, but inherits some of the same shortcomings as mean-field theories. Alternatively, if the assumption of spherical particles is valid, then a closed form for Greens' function, in the form of a harmonic sum, can be obtained in some cases.\cite{gan2015} Moreover, a closed form solution for Greens' function can sometimes be obtained using the Image Charge Method (ICM) for simple geometries.\cite{gan2015,xu_fast_2011,xu_image_2011} Other standard methods that do not require the discretization of the simulated volume are the Method of Moments (MoM)\cite{Sangani1983,zangwill2013} and the perturbative many-body expansion method,\cite{freed_perturbative_2014} which finds the potential (including polarization) from a scattering expansion. These techniques, however, can be prohibitively expensive if large systems are studied (e.g. ICM scales with the number of particles, $N$, as $\sim N^3$, for large $N$) or the particles are too close together (for example, MoM requires a large number of moments to converge in concentrated systems). Hybrid methods have also emerged to simulate spherical particles that scale linearly with the number of particles in the system\cite{gan_efficient_2019}, but are strictly applicable to colloidal spherical particles. Significant effort has been put into decreasing the computational time required to solve for the induced charge effects in soft matter systems relying on boundary element formulations.\cite{barros_dielectric_2014,jiang_2016,nguyen_incorporating_2019}
For example, the induced charge has been related to the electric field to obtain a system of equations in the form of $A \mbf{s} = \mbf{b} $, where $\mbf{b}$ depends on the free charge, $\mbf{s}$ is a vector of discretization points on the surface of interest. $A$ depends on the geometry of the system and can take the form of a linear operator $\mathcal{A}$,\cite{barros_dielectric_2014} or as weights of a linear combination of nodes, $\mbf{A}$.\cite{tyagi_iterative_2010}  The $\mrm{ICC}^{\star}$ method, for example, proposes to use the successive over-relaxation method to solve for the charge distribution at a later time step in coarse-grained dynamic simulations, avoiding the need to invert the weighting coefficients matrix each time step.\cite{tyagi_iterative_2010} \\
\indent  When solving for the electrostatic potential, an additional obstacle arises due to the long-range nature of the electrostatic forces, which decay as $O(1/r)$ with the distance from the particle, $r$. This becomes particularly demanding when periodic boundary conditions are used. In particle-based computer simulations, these are commonly computed using mesh-based variants of the Ewald sum method, which splits the interparticle interactions into short and long range and solves for them in real and Fourier space, respectively.\cite{deserno_how_1998,nguyen_incorporating_2019, arnold_electrostatics_2002} For non-periodic systems, the Fast Multipole Method is preferred. \cite{poursina_improved_2014,greengard_fast_1997} \\
%
\indent Given the limitations of non-mesh-based methods for simulating systems with a large number of particles and the inability of Poisson-Boltzmann-type treatments to accurately consider packing and polarization effects, it is necessary to develop even faster methods for dynamic simulations of colloidal systems. Here, we apply the immersed boundary method
to simulate the electrostatically stabilized colloidal suspensions at the mesolevel. We take advantage of an spectrally accurate Ewald summation type (SE, for spectral Ewald) driven by graphics processing units (GPUs) to accelerate the computation of the electrostatic forces and polarization due to local and external fields. The SE technique has also been used to calculate long-range interparticle hydrodynamic interactions for dynamic simulations of dense suspensions, \cite{fiore_swan_2019, wang_spectral_2016} and, more recently, for the rapid computation of the stress tensor of mutually polarizable spheres.\cite{reed_stress_2021} The manuscript is organized as follows, Sect. \ref{sec:background},  brief background on the formulation of electrostatic interactions between spherical particles in a periodic geometry. In Sect. \ref{sec:IMBOUND}, we describe the immersed boundary method to calculate electrostatic forces on arbitrarily shaped polarizable bodies. Section \ref{sec:StaticSims} presents a series of validation calculations for the immersed boundary method. Finally, Sect. \ref{sec:BDofColloids} shows its use for Brownian dynamic simulations of electric double layers and the response of an ideally polarizable colloidal particle immersed in an electrolyte to an external electric field. These out-of-equilibrium simulations show the applicability of the method to obtain dynamic properties of colloidal suspensions, such as the electrophoretic velocity of charged colloids as a result of an electric field and the induced charge electro-osmosis phenomenon. \cite{squires2004} \\
\section{ \label{sec:background} Background / Grand Potential Matrix derivation}
Electrostatic phenomena in colloidal dispersions are described by Poisson's equation for the electrostatic potential $\psi(\mathbf{x})$
 \begin{equation}
     \nabla \cdot \mbf{E} = \frac{\rho}{\varepsilon} \;,\quad \mbf{E}=-\nabla \psi
 \end{equation}
 where $\mbf{E}(\mathbf{x})$ is the electric field, $\rho(\mbf{x})$ is the free charge density and $\varepsilon(\mathbf{x})$ is the electric permittivity. 
 %
 The free charge is localized to colloids and ions so that $\rho(\mathbf{x})=0$ everywhere for $\mathbf{x}$ in the fluid. These assumptions reduce Poisson's equation to the simpler Laplace equation with boundary conditions on the surfaces,
\begin{equation}
    \nabla^2\psi=0\:,
    \label{eq:laplace}
\end{equation}
\begin{equation}
    \psi_{\mathrm{p}} = \psi_{\mathrm{f}} , \hspace{4mm} (\varepsilon_\mrm{f} \mathbf{E}_{\mathrm{f}} - \varepsilon_\mrm{p} \mathbf{E}_{\mathrm{p}}) \cdot \hat{\mathbf{n}} = q_{\alpha}/(4 \pi a^2)\:, \hspace{2mm}
    \label{eq:BCs}
\end{equation}
where ${\psi_{\mathrm{p}}}$ and ${\psi_{\mathrm{f}}}$ are the potentials inside and outside the particle, ${q_{\alpha}}/(4\pi a^2)$ is the uniform free surface charge density of ion $ \alpha $ on a spherical shell with radius $ a $ and net charge ${q_{\alpha}}$,  $\mathbf{\hat{n}}$ is the normal outward vector and ${\mathbf{E}_{\mathrm{f}} }$ and ${\mathbf{E}_{\mathrm{p}} }$ are the electric field outside and inside the particle, respectively. \cite{Jackson1999, Landau1984} Further, there may be an externally applied electric field and potential $\mathbf{E}_0 = \nabla \psi_0$, where $\psi(\mathbf{x}) = \psi_0$ in the absence of particles, and we require triply periodic boundary conditions. The time scale of particle motion, $O(\SI{1}{\mu m}$--$\SI{1}{s})$, is orders of magnitude greater than the electric and magnetic relaxation times at the atomic level, $O(10^{-9}\mrm{s})$, the electric potential $\psi(\mathbf{x})$ is pseudo-steady, and time dependence emerges solely through the time-varying boundary conditions in \eqref{eq:BCs}.
For a point ${\mathbf{x}}$ in the fluid, the potential is given by the integral form of Laplace's equation;
\begin{align}\label{eq:psi_f_1}
   \psi_{\mathrm{f}}(\mathbf{x})-\psi_{0}(\mathbf{x}) &= \frac{1}{\varepsilon_{\mrm{f}}} \sum_{\alpha} \int_{S_{\alpha}} \mrm{d} \mathbf{x}^{\prime} \Biggl( G(\mathbf{x}-\mathbf{x}^{\prime}) \mathbf{E}_{\mathrm{f}} (\mathbf{x}^{\prime}) 
   \cdot\hat{\mathbf{n}}_{\mathbf{x}^{\prime}} \\  \nonumber
   &+ \varepsilon_{\mathrm{f}}\psi_{\mathrm{f}}(\mathbf{x}^{\prime})\hat{\mathbf{n}}_{\mathbf{x}^{\prime}}\cdot\nabla_{\mathbf{x}^{\prime}}G(\mathbf{x}-\mathbf{x}^{\prime}) \Biggr) \; ,
\end{align}
where $G$ is Green's function for Laplace's equation. We can enforce periodic boundary conditions on $\psi$ at this stage by using the periodic Green's function satisfying $-\nabla^2G(\mathbf{x}) = \sum_\mathbf{n} \delta(\mathbf{x}-\mathbf{n})$ where $\mathbf{n} = [n_xL_x, n_yL_y, n_zL_z]$, $n_i$ are integers, and $L_i$ are the periodicities in each dimension. This periodic Green's function admits a spherical harmonic expansion;
\begin{align}
&G(\mathbf{x}-\mathbf{y}) = \\ \nonumber
&\frac{1}{V} \sum_{\mathbf{k} \neq 0} \frac{e^{i \mathbf{k} \cdot \mathbf{x}}}{k^2} = \frac{1}{V} \sum_{\mathbf{k} \neq 0} \frac{e^{i \mathbf{k} \cdot (\mathbf{x} - \mathbf{x}_j)}}{k^2} \sum_\ell \sqrt{4\pi(2\ell + 1)}i^{-\ell}\\  \nonumber
&\times j_\ell(ka) \left( \frac{r}{a} \right)^\ell Y_\ell(\theta,\phi)\;,
\end{align}
which allows \eqref{eq:psi_f_1} to be written in terms of multipole moments,
\begin{align}
    &\psi_{\mathrm{f}}(\mathbf{x})-\psi_{0}(\mathbf{x})=\\ \nonumber
    &\frac{\varepsilon_{\mathrm{f}}}{V}\sum_{\mathbf{k}\neq0}\sum_{\alpha}\frac{e^{i\mathbf{k}\cdot(\mathbf{x}-\mathbf{x}_{\alpha})}}{k^{2}} \left( q_{\alpha}j_{0}(ka) - \frac{ 3 i }{ a } j_{1}(ka) \mbf{S}_{\alpha} \cdot \hat{ \mbf{k} } \cdots \right) \;,
    \label{eq:psi_f}
\end{align}
where ${j_{i}(x)}$ is the spherical Bessel function of the first kind of $i^{\mrm{th}}$ order, $\mathbf{k} = [2\pi \kappa_x/L_x, 2\pi \kappa_y/L_y, 2\pi \kappa_z/L_z]$ for integers $\kappa_i$, and $\hat{\mbf{k}}$ the unit vector in the $\mbf{k}$ direction. The $j_{i}(ka)$ terms are called ``shape factors'' that account for how spherical particles propagate electric potentials through Fourier space. Note that due to periodic boundary conditions, equation \eqref{eq:psi_f_1} should also contain a contribution from an integral over a closed surface at infinity ($S_\infty$). Bonnecaze and Brady show\cite{Bonnecaze1990} that including this $S_\infty$ integral is equivalent to enforcing electroneutrality by removing the $\mathbf{k}=0$ term, so we will not consider it here. \cite{Sherman2019}
Integrating on the surface of each particle, applying the boundary conditions in \eqref{eq:BCs}, and retaining only the first moment of expansion,
\begin{equation}
\psi_\alpha(\mathbf{x}_\alpha) - \psi_0(\mathbf{x}_\alpha) = \frac{\varepsilon_{\mathrm{f}}}{V}\sum_{\mathbf{k}\neq0} \sum_\beta j_{0}^{2}(ka)\frac{e^{i\mathbf{k}\cdot(\mathbf{x}_{\alpha}-\mathbf{x}_{\beta})}}{k^{2}} q_\beta \; .
\label{eq:pot_charge}
\end{equation}
where $\psi_\alpha = \frac{1}{4 \pi a^2 } \int_{S_{\alpha}} \mrm{d} \mbf{x}  \psi_{\mrm{p}}(\mbf{x})$ is the surface average $\psi_\mrm{p}$ of the particle $\alpha$. Note that the $\beta$ sum \emph{includes} the $\beta=\alpha$ term, i.e. the self-term, is naturally incorporated into \eqref{eq:M00_Fourier}. Higher-order moments can be retained in \eqref{eq:M00_Fourier},\cite{Sherman2019} e.g. to account for dipole-dipole interactions among particles,\cite{Sherman2018} but here we consider only charges. Equation \eqref{eq:M00_Fourier} is a system of equations that relates each of the $N$ particle potentials $\psi_\alpha$ to the $N$ particle charges $q_\beta$, which we represent in matrix-vector form,
\begin{equation} \label{eq:Moments}
\Psi -\Psi_{0}=\mathcal{M}^{\mrm{E}}\cdot\mathcal{Q}\;,
\end{equation}
where $\Psi -\Psi_{0}=[\psi_{1} -\psi_{0}(\mathbf{x}_{1}), \psi_{2} -\psi_{0}(\mathbf{x}_{2}),...,\psi_{N} -\psi_{0}(\mathbf{x}_{N})]^{T}$
is a list of particle potentials relative to the external potential at their centers, $\mathcal{Q} =[q_{1},q_{2},...,q_{N}]^T$ is the list of particle charges, and $\mathcal{M}^{E}$ is named the potential tensor, with elements
\begin{equation}
M_{\alpha\beta} = \frac{\varepsilon_{\mathrm{f}}}{V}\sum_{\mathbf{k}\neq0}j_{0}^{2}(ka)\frac{e^{i\mathbf{k}\cdot(\mathbf{x}_{\alpha}-\mathbf{x}_{\beta})}}{k^{2}} \; .
\label{eq:M00_Fourier}
\end{equation}
In real space, these elements are
\begin{equation}
M_{\alpha\beta} = \begin{dcases} \frac{1}{4\pi a \varepsilon_f r}, & r \geq 2a \\
\frac{4a-r}{16\pi a \varepsilon_f}, & r < 2a
\end{dcases}
\end{equation}
$M_{\alpha \beta}$ is identical to the Coulomb interaction when the particles do not overlap ($r \geq 2a$). For overlapping particles ($r < 2a$), $M_{\alpha \beta}$ regularizes this Coulomb interaction. Though overlaps do not correspond to any physically-realizable configuration, they often occur in many kinds of molecular simulations. It is impossible to satisfy the boundary conditions \eqref{eq:BCs} for overlapping particles, so the choice of regularization is arbitrary. The particular form of $M_{\alpha \beta}$ here is advantageous because (1) it ensures that $\mathcal{M}^E$ is always positive definite, (2) it ensures that $\mathcal{M}^E$ is well-conditioned, as no elements diverge, and (3) it decays more quickly $(\sim 1/k^4)$ than normal Coulomb interactions $(\sim 1/k^2)$. As we explain in Section \ref{sec:Ewald}, these features enable efficient numerical calculations to compute electrostatic forces.
\section{\label{sec:IMBOUND}Immersed Boundary method for electric forces of arbitrarily shaped conductors }
%
Immersed boundary (IB) methods for solving partial differential equations provide a faster alternative to standard body-conforming gridding methods for dealing with moving boundaries and interfaces with complex geometries. In general, immersed boundaries require two independent discretizations: one discretization for the body surfaces and a second for the fluid volume. \\
IB methods have been useful for simulating fluid flow and hydrodynamic interactions in colloidal dispersions,\cite{SwanGang2016} electrohydrodynamics\cite{HU2015} and flow near elastic boundaries\cite{peskin_flow_1972}. Here, we derive an immersed boundary method for colloidal particles of \emph{arbitrary shape and charge}. The method facilitates the computation of induced polarization on the surface of complex geometries due to local (by other charged bodies) and external fields.%
\subsection{\label{sec:Composite_construction}Composite Bead Model}
\indent In general, we propose to represent the bodies with a composite-bead structure where beads tessellate the bodies' surfaces. Although these beads can take any size, it is convenient to use a single geometric length scale. In our particle-based simulations, the natural and shortest geometric length scale is given by the ion's diameter. Thus, we constrain the tessellating beads to the same diameter in value ${2a}$. 

\indent Simple spherical bodies can be constructed iteratively by placing $12$ beads on the vertices of an icosahedron. Then, we can bisect the edges with more beads; projecting the new beads radially outward such that their centers lie on the body surface. This procedure can be repeated until a user-defined discretization level is reached,\cite{Swan2016} This process builds a series of ``icospheres'' that provide increasing resolution of the body surface as the number of composing beads increases. We have previously found that setting the beads to be as large as possible without overlapping yields a good approximation of the true surface.\cite{SwanGang2016}  This approach is a type of ``immersed boundary method'' because we use different, incommensurate discretizations of the body surfaces and the fluid. Surfaces are discretized with the bead tessellation and, consequently, are ``immersed'' in the Fourier grid that discretizes the fluid.\cite{Sherman2019} The error introduced by the level of refinement is tested in sec. \ref{sec:StaticSims}.\\
\subsection{ Induced Surface Charge Distribution of Polarizable Bodies}
%
%
%
\indent Consider a suspension with $N_b$ bodies, where the center of mass of the ${i^\mathrm{th}}$ body is $\mathbf{X}_{i}$. The surface of each body is made up of $N_{i}$ beads of radius $a$ at positions $\mathbf{x}_{i j}$. Ionic species can be considered to be bodies with one constituent particle. Note that we use lowercase variables to indicate individual bead quantities and uppercase variables to indicate rigid body quantities. Lowercase bead quantities with two indices ($i j$) indicate the $j^\text{th}$ bead on the $i^\text{th}$ body, while a single index ($\beta$) implies a global bead index (\emph{i.e.} $\beta = (i-1)N_i + j$). The net charge on the $i^{\text{th}}$ body is equal to the sum of its constituent bead charges $q_{i j}$, 
\begin{equation}
Q_{i} = \sum_{j} q_{i j}\:.
\label{eq:Q_tot}
\end{equation}
Here, we assume that bodies are ideally polarizable (i.e. $\varepsilon_\mrm{p} \rightarrow \infty$), so that the electric potential everywhere in the body is constant and that each bead's potential $\psi_{i j }(\mathbf{x}_{i j})$ is equal to the body's potential $\Psi_{i}(\mathbf{X}_{i})$ to which it belongs, $\psi_{i j } = \Psi_{i}$. \\
\indent Bodies with finite $\varepsilon_\mrm{p}$ have a more sophisticated relation between bead potential and body potential\cite{barros_dielectric_2014} and are the subject of future study. The suspension is immersed in a constant electric field, $\mathbf{E}_0$, which establishes the external potential $\psi_0(\mathbf{x}) = - \mathbf{x} \cdot \mathbf{E}_0$. The difference between the bead potential and the external potential is
\begin{equation} \label{eq:beadpot1}
\psi_{i j } - \psi_{0,i j } = \Psi_{i} - \Psi_{0,i} + \mathbf{r}_{i j } \cdot \mathbf{E}_0,
\end{equation} 
where $\mathbf{r}_{i j} \equiv \mathbf{x}_{i j} - \mathbf{X}_{i}$ is the position of the bead relative to the center of mass of the body. The bead potentials $\phi_\alpha$ are also related to the body charges $q_\beta$ through the potential tensor $\mathbf{M}$ in \eqref{eq:pot_charge}. 

\indent Typically, the body charges $Q_{i}$ and the external field $\mathbf{E}_0$ are known, while the induced bead charges $q_\alpha$ and the body potentials $\Psi_{i}$ are unknown. We can construct a system of equations for these unknowns by combining equations \eqref{eq:pot_charge}, \eqref{eq:Q_tot}, and \eqref{eq:beadpot1}
\begin{equation} \label{eq:saddle}
\begin{bmatrix}
-\mathbf{M} & \boldsymbol{\Sigma}^T \\
\boldsymbol{\Sigma} & 0
\end{bmatrix} \cdot \begin{bmatrix}
\mathbf{q} \\
\boldsymbol{\Psi} - \boldsymbol{\Psi}^0
\end{bmatrix} = \begin{bmatrix}
-\mathbf{r} \cdot \mathbf{E}^0 \\
\mathbf{Q}
\end{bmatrix}.
\end{equation}
$\mathbf{q} \equiv [q_{11}, q_{12}, \cdots, q_{21}, q_{22}, \cdots]^T$ is a list of all $N$ bead charges, $\mathbf{Q} \equiv [Q_1, Q_2, \cdots]^T$ is a list of all $N_b$ body charges, $\boldsymbol{\Psi}-\boldsymbol{\Psi}_0 \equiv [\Psi_1-\Psi_{0,1}, \Psi_2-\Psi_{0,2}, \cdots]^T$ is a list of the difference between the $N_b$ body potentials and the external potential, and $\mathbf{r} \cdot \mathbf{E}_0 \equiv [\mathbf{r}_{11}\cdot \mathbf{E}_0, \mathbf{r}_{12}\cdot \mathbf{E}_0, \cdots, \mathbf{r}_{21}\cdot \mathbf{E}_0, \mathbf{r}_{22}\cdot \mathbf{E}_0, \cdots]^T$ is a list of $N$ relative bead positions dotted with the external field. $\boldsymbol{\Sigma}$ is an $N_b\times N$ summation tensor whose rows correspond to bodies and columns correspond to beads. Each row is entirely $0$, except for $N_{i}$ consecutive $1$ values in the columns corresponding to the $N_{i}$ beads of the $i^{\mrm{th}}$ body. Further details for constructing $\boldsymbol{\Sigma}$ and subsequent summation tensors are found in the Appendix. Eq. \ref{eq:saddle} can be solved iteratively\cite{Barros2014} for the bead charges of and the rigid body potentials. \\
\indent The induced dipole moment of a rigid body can then be computed from the bead charges
\begin{equation}
\mathbf{S}_{i} = \sum_{j} \mathbf{r}_{ i j } q_{ i j }.
\label{eq:rigid_dipole}
\end{equation}
 Note that for single-bead bodies $\mathbf{r}_{ i j }=0$. The set of all rigid body moments (charges and dipoles) is then given by
\begin{equation}
\begin{bmatrix}
\mathbf{Q} \\
\mathbf{S}
\end{bmatrix} = \boldsymbol{\Sigma}' \cdot \mathbf{M}^{-1} \cdot \boldsymbol{\Sigma}'^T \cdot \begin{bmatrix}
\boldsymbol{\Psi}-\boldsymbol{\Psi}_0 \\
\mathbf{E}_0
\end{bmatrix}.
\label{eq:QS_capacitance}
\end{equation}
where $\boldsymbol{\Sigma}'$ is an $4N_b\times N$ summation tensor, whose first $N_b$ rows are identical to $\boldsymbol{\Sigma}$ and whose next $3N_b$ rows look structurally like $\boldsymbol{\Sigma}$ but with each $1$ value replaced with $\mathbf{r}_{ij}$ (see Appendix). The quantity $\mbf{C} = \boldsymbol{\Sigma}' \cdot \mathbf{M}^{-1} \boldsymbol{\Sigma}'^T$ is called the ``grand capacitance tensor'' of the dispersion and will be referenced in sec. \ref{sec:ArrayConductivity}.

\subsection{ Forces and Torques }

The potential energy of the suspension is expressed as the sum of products of rigid body moments and potential derivatives \cite{Jackson1999}
\begin{align} \label{eq:pot_syst}
U &= \frac{1}{2} \sum_{i} \left( Q_{i} \Psi_{i} - \mathbf{S}_{i} \cdot \mathbf{E}_0 \right) \\ \nonumber
&=\frac{1}{2}\mbf{Q}\cdot(\left\langle \boldsymbol{\Psi}\right\rangle -\boldsymbol{\Psi}_{0})\\ \nonumber
&=\frac{1}{2}\mbf{Q}\cdot\boldsymbol{\Psi}_{0} + \frac{1}{2} \left[ \mathbf{q} \quad \left\langle \boldsymbol{\Psi}\right\rangle -\boldsymbol{\Psi}_{0} \right] \cdot \begin{bmatrix}
-\mathbf{r} \cdot \mathbf{E}_0 \\
\mathbf{Q} 
\end{bmatrix}\;.
\end{align}
\indent Moreover, the force on each bead is the negative derivative of $U$ with respect to the bead position $\mathbf{x}_\alpha$, given all other beads remain fixed, $\mbf{f}_\alpha = -\nabla_{\mathbf{x}_\alpha} U$. Using \eqref{eq:saddle} in \eqref{eq:pot_syst} and taking the gradient (see Appendix), the force on bead $\alpha$ is
\begin{equation} \label{eq:beadforce}
\mathbf{f}_\alpha = q_\alpha \mathbf{E}_0 - \frac{1}{2} \left( \nabla_{\mathbf{x}_\alpha} \mathbf{M} \right) : \mathbf{q}\mathbf{q} 
\end{equation}
From the distribution of bead forces, we can compute the $i^{\mrm{th}}$ body's force, $\mathbf{F}_{i} $ and torque $\mathbf{L}_{i}$
\begin{equation}
   \mathbf{F}_{i} = \sum_{j} \mathbf{f}_{i j }, \quad \mathbf{L}_{i}  = \sum_{j} \mathbf{r}_{i j } \times \mathbf{f}_{i j }.
\end{equation}
These forces and torques serve as the input to a rigid body hydrodynamic integrator\cite{Swan2016} that computes the rigid body velocities and angular velocities of the composites.
\subsection{\label{sec:Ewald} Spectrally-accurate Ewald summation}
%
%
The speed at which we can iteratively solve \eqref{eq:saddle} is limited by the speed at which we compute the potential-dipole sums in \eqref{eq:pot_charge}. Here we use a particularly efficient, spectrally accurate Ewald summation effort from ref. \citenum{Lindbo2011} which we adapt for our regularized form of the Coulombic interaction \eqref{eq:M00_Fourier} following ref. \cite{Swan2016}. The sum in \eqref{eq:pot_charge} is split into two rapidly convergent sums
\begin{align}\label{eq:splitting}
    &\left\langle \psi\right\rangle _{\alpha}-\psi_{0}(\mathbf{x}_{\alpha})=\frac{\varepsilon_{\mathrm{f}}}{V}\sum_{\mathbf{k}\neq0}\sum_{\beta}\frac{e^{i\mathbf{k}\cdot(\mathbf{x}_{\alpha}-\mathbf{x}_{\beta})}}{k^{2}}h(k)j_{0}^{2}(ka)\cdot q_{\beta} \\ \nonumber
&+\varepsilon_{\mathrm{f}}\sum_{\mathbf{n}}\sum_{\beta}\mathcal{F}^{-1}\left\{ \frac{e^{i\mathbf{k}\cdot(\mathbf{x}_{\alpha}-\mathbf{x}_{\beta})}}{k^{2}}\left(1-h(k)\right)j_{0}^{2}(ka)\right\} \cdot q_{\beta},
\end{align}
 where ${h(k)\equiv\exp(-k^2 / 4\xi^2 )}$, ${\xi}$ is the Ewald splitting parameter controlling the convergence rate of each sum, and ${\mathcal{F}^{-1}}$ is the inverse Fourier transform.
\subsubsection{\label{sec:ReEwald} Real-space sum}
The real-space contribution to \eqref{eq:splitting} can be computed by explicitly performing the inverse Fourier transform and then directly summing,
\begin{widetext}
\begin{align} \label{eq:psi_Re_1}
\left\langle \psi\right\rangle _{\alpha}^r-\psi_{0}^{r}(\mathbf{x}_{\alpha}) &=\frac{1}{\varepsilon_f} \sum_\beta q_\beta \bigg( -\frac{r+2a}{32\pi^{3/2}a^2 \xi r} e^{-(r_{\alpha \beta }+2a)^2 \xi^2} -\frac{r-2a}{32\pi^{3/2}a^2 \xi r} e^{-(r_{\alpha\beta}-2a)^2 \xi^2} + \frac{1}{16\pi^{3/2}a^2 \xi} e^{-r_{\alpha\beta}^2 \xi^2}\\ \nonumber
&+ \frac{2\xi^2 (r + 2a)^2 + 1}{64 \pi a^2 \xi^2 r} \erfc (r_{\alpha\beta}+2a)\xi + \frac{2\xi^2 (r - 2a)^2 + 1}{64 \pi a^2 \xi^2 r}\erfc \left( r_{\alpha\beta}-2a \right) \xi \\ \nonumber
&-\frac{2 \xi^2 r^2+1}{32\pi a^2\xi^2r}\erfc r_{\alpha\beta}\xi - \left(\frac{1}{4\pi r}+ \frac{4a-r}{16\pi a^2}\right) \mathrm{H}(2a - r_{\alpha\beta}) \bigg)\:, \nonumber
\end{align}
\end{widetext}
where ${\mathrm{H}}$ is the Heaviside step function and $\erfc$ is the complementary error function. Note that the Heaviside term emerges naturally from the inverse Fourier transform and ``turns on'' the regularization in \eqref{eq:M00_Fourier} for overlapping particles (in fact, this is how the regularization was obtained).  The ``self-term'' also emerges naturally from the ${r\rightarrow 0}$ limit,
\begin{equation}
\left\langle \psi\right\rangle _{\alpha}^{\mathrm{s}}-\psi_{0}^{\mathrm{s}}(\mathbf{x}_{\alpha}) = \frac{1}{\varepsilon_f}\left( \frac{1-e^{-4a^2\xi^2}}{8\pi^{3/2} a^2 \xi} + \frac{ \erfc 2a\xi}{4\pi a} \right) q_\alpha.
\end{equation}
Note that the point charge result\cite{Lindbo2011} is recovered by letting ${a\rightarrow0}$,
\begin{equation}\label{eq:psi_Re_point}
\left\langle \psi\right\rangle _{\alpha}^{\mathrm{Re}}-\psi_{0}^{\mathrm{Re}}(\mathbf{x}_{\alpha})= \frac{1}{4\pi\varepsilon_f} \sum_\beta q_\beta \frac{\erfc{\xi r_{\alpha\beta}}}{r_{\alpha\beta}}.
\end{equation}
The sum of the real space decays exponentially with the distance and can converge rapidly by controlling $\xi$ up to a user-defined cutoff radius, $r_\mrm{c}$. This truncation leads to an associated truncation error, $\bar{\delta}_{\mrm{r}}$, whose bounds can be approximated as \cite{Kolafa1992}
\begin{equation} \label{eq:realspaceerror}
\bar{\delta}_{\mrm{r}}\leq C e^{-r_c^2 \xi^2},
\end{equation}
where $C$ is some constant. The sum of the local interactions on the GPU is performed by assigning one thread per particle and looping through their neighbors. The neighbor list can be handled using a linear complexity linked cell algorithm\cite{allen_computer_2017} which divides the simulation box into cells of size $r_{\mrm{c}}\sim \xi^{-1}$ and considers interactions only between particles belonging to the same or adjacent cells.\cite{Fiore2018}
\subsubsection{\label{sec:WEwald} Wave-space sum}

 The wave space contribution to \eqref{eq:splitting} is computed with the spectral Ewald method proposed by Lindbo and Tornberg,\cite{Lindbo2011} which introduces a parameter ${\eta}$ to split the Ewald splitting function,${h(k)=\exp(-k^2/4\xi^2)}$, even further, 
\begin{align}
&\frac{\varepsilon_{\mathrm{f}}}{V}\sum_{\mathbf{k}\neq0}\frac{e^{i\mathbf{k}\cdot\mathbf{x}_{\alpha}}}{k^{2}}e^{-k^{2}/4\xi^{2}}\sum_{\beta}q_{\beta}e^{-i\mathbf{k}\cdot\mathbf{x}_{\beta}}j_{0}^{2}(ka)\\ \nonumber
&=\frac{\varepsilon_{\mathrm{f}}}{V}\sum_{\mathbf{k}\neq0}\frac{e^{i\mathbf{k}\cdot\mathbf{x}_{\alpha}}}{k^{2}}e^{-(1-\eta)k^{2}/4\xi^{2}}j_{0}^{2}(ka)\left(\sum_{\beta}q_{\beta}e^{-i\mathbf{k}\cdot\mathbf{x}_{\beta}}e^{-\eta k^{2}/4\xi^{2}}\right)\;,
\label{eq:splitting2}
\end{align}
and takes advantage of the fact that the Fourier transform of a Gaussian is a Gaussian, and the shape of this Gaussian is controlled by $\eta$. In particular, for the term in parentheses in \eqref{eq:splitting2} 
\begin{align} 
   \mathbf{\widehat{H}}(\mathbf{k}) &\equiv\sum_{\beta}q_{\beta}e^{-\eta k^{2}/8\xi^{2}}e^{-i\mathbf{k}\cdot\mathbf{x}_{\beta}}, \\ \nonumber
   \mathbf{H}(\mathbf{x}) &=\left(\frac{2\xi}{\pi\eta}^{2}\right)^{3/2}\sum_{\beta}q_{\beta}e^{-2\xi^{2}(\mathbf{x}-\mathbf{x}_{\beta})^{2}/\eta}
   \label{eq:H_hat}, 
\end{align}
 After defining a ``scaled'' $\widehat{\widetilde{\mathbf{H}}}$,
\begin{equation}
\widehat{\widetilde{\mathbf{H}}}(\mathbf{k})\equiv j_{0}^{2}(ka)\cdot\frac{e^{-(1-\eta)k^{2}/4\xi^{2}}}{k^{2}}\widehat{\mathbf{H}}(\mathbf{k}),
\end{equation}
we use Parseval's formula to compute the rest of the wave-space sum,
\begin{align} 
    \left\langle \psi\right\rangle _{\alpha}^k-\psi_{0}^k(\mathbf{x}_{\alpha}) &=\frac{1}{\varepsilon_{\mathrm{f}}V}\sum_{\mathbf{k}\neq0}\widehat{\widetilde{\mathbf{H}}}_{\mathbf{k}}e^{i\mathbf{k}\cdot\mathbf{x}_{\alpha}} \\ \nonumber
    &=\frac{1}{\varepsilon_{\mathrm{f}}}\left(\frac{2\xi}{\pi\eta}^{2}\right)^{3/2}\int_{V}d\mathbf{x}\mathbf{\widetilde{H}}(\mathbf{x})e^{-2\xi^{2}(\mathbf{x}-\mathbf{x}_{\alpha})^{2}/\eta}\;,
    \label{eq:psi_wave}
\end{align}
where ${\mathbf{\widetilde{H}}}$ is the inverse transform of ${\widehat{\mathbf{\widetilde{H}}}}$.\\
\indent As in the real space sum (see Sect. \ref{sec:ReEwald}), there is a truncation error for the wave-space sum, $\bar{\delta}_{\mrm{k}} $, whose bounds can be approximated as
\begin{equation}
    \bar{\delta}_{\mrm{k}} \leq C e^{-k_c^2/4\xi^2},
\end{equation}
where $C$ is some constant. In our simulations, $k_c\geq |
\mbf{k}|$ corresponds to a real/wave space cubic grid of $N_g = 1 + Lk_c/\pi$ nodes in each dimension with spacing $h \equiv L/N_g$. This regular grid is needed to evaluate fast Fourier transforms (FFTs) and inverse fast Fourier transforms (IFFTs).\\%
As particles' positions do not tend to perfectly coincide with the points in a regular grid, we need to map the potential and potential gradients from the particle's positions to nearby points on the grid. This is the role of $\mbf{H}(\mbf{x})$, in eq. \ref{eq:H_hat}, which uses Gaussians to ``spread'' or project the particle's moments on these regular grid points. \\
\indent The number of grid points over which we extrapolate the information (the potential and its gradients) can be significantly reduced by noticing that ${\mbf{H}(\mbf{x})}$ also decays exponentially. Consequently, we can truncate the spread to a cubic array of the closest $P$ grid points in each dimension in $O(NP^3)$ calculations, resulting in the truncation error
\begin{equation} \label{eq:truncerror}
\bar{\delta}_\mrm{t} \leq C e^{-h^2 P^2 \xi^2/2\eta}\;.
\end{equation}
\indent Similarly, once $\mbf{\tilde{\mbf{H}}(\mbf{x})}$ is computed at the grid points via the IFFT of $\Hat{\tilde{\mbf{H}}}(\mbf{x})$, we interpolate the information from the regular grid points to the particle positions by integrating eq. \ref{eq:psi_wave} numerically using the trapezoid rule.\cite{numerical_2007} For periodic boundary conditions, the trapezoidal rule quadrature is spectrally accurate\cite{Lindbo2011} with a quadrature error given by\cite{Lindbo2011}
\begin{equation}
\bar{\delta}_\mrm{q} \leq C e^{-\pi^2 \eta/2 \xi^2 h^2}.
\end{equation}
\indent As FFT and IFFT are accurate up to machine precision,\cite{numerical_2007,WangBrady2016}, the parameters $\eta$ and support grid points, $P$, collectively control the numerical error introduced by the spreading and contracting steps.
The natural choice of $\eta$ \cite{Lindbo2010} is obtained by forcing the resulting error from truncating the spreading grid points, $\bar{\delta}_{\mrm{t}}$, to be equal to the quadrature error, $\bar{\delta}_{\mrm{q}}$. Thus, we find a simple relation between $\eta$ and $P$,
\begin{equation}
    \eta = \frac{h^2 \xi^2 P }{ \pi }\;.
\end{equation}
This choice of $\eta$ ensures that the Gaussian width is wide enough so that the convolution can be easily approximated without incurring unnecessary truncation errors because the Gaussian kernel is not narrow enough. \\
The error for the entire contraction step can be expressed as
\begin{equation} \label{eq:contractionerror}
\bar{\delta}_{\mrm{c}} \leq C e^{-\pi P/2}.
\end{equation}
\subsubsection{ Spectral Ewald implementation }
\indent The following algorithm can be used to compute the particle's potential in eq. \ref{eq:splitting} and potential gradients in each time step. Here, we provide additional details of the GPU implementation of our method (see Supporting Material).\\
\indent \textbf{1. Real Space Sum.}  
The sum of the local interactions on the GPU is carried out by assigning one thread per particle and looping through their neighbors. The neighbor's list can be handled using a linear complexity linked cell algorithm which divides the simulation box into cells of size $r_{\mrm{c}}\sim \xi^{-1}$ and considers interactions only between particles belonging to the same or adjacent cells.\cite{Fiore2018}

\indent \textbf{2. Spreading.}  
Divide the simulation box into a Cartesian cubic grid (or over the rectangular domain of the simulation box when the dimensions are unequal) and compute ${\mbf{H}(\mbf{x})}$ using eq. \ref{eq:H_hat} on the grid points.On the GPU, spreading is performed using one block of $P^3$ threads per particle, with each thread corresponding to one of the $P^3$ supporting grid points.  

\indent \textbf{3. Fast Fourier Transform} FFTs are computed for $\mathbf{H}(\mathbf{x})$ in each Cartesian direction to obtain $\widehat{\mathbf{H}}(\mathbf{k})$. Efficient algorithms, such as the NVIDIA cuFFT library, \cite{web:nvidia}, can perform this operation with $O(N_g^3 \log N_g^3)$ complexity.

\indent \textbf{4. Scaling.} We convert particle moments into potentials in wavespace by scaling $\widehat{\mathbf{H}}(\mathbf{k})$ at each of the grid points. Computing $\widehat{\widetilde{\mathbf{H}}}(\mathbf{k})$ requires $O(N_g^3)$ operation and a single execution thread is assigned to each point in the grid. 

\indent \textbf{5. Inverse Fast Fourier Transform.}  As with the FFT step, we compute $\widetilde{\mathbf{H}}(\mathbf{x})$ on the grid points. This step requires $O(N_g^3 \log N_g^3)$ operations.

\indent \textbf{6. Contraction.}  We interpolate $ \widetilde{ \mbf{H} }(\mbf{x}) $ from the equally spaced grid points to the particles' positions using Gaussians to arrive to \ref{eq:psi_wave}, evaluated numerically using quadrature \emph{via} the trapezoidal rule. Finally, we add the contributions to the potential and its gradients from the real and imaginary parts.

\section{\label{sec:StaticSims} Polarizable colloids in an external electric field}
%
\begin{figure*}[!t]
\centering
\includegraphics[width=0.9\textwidth]{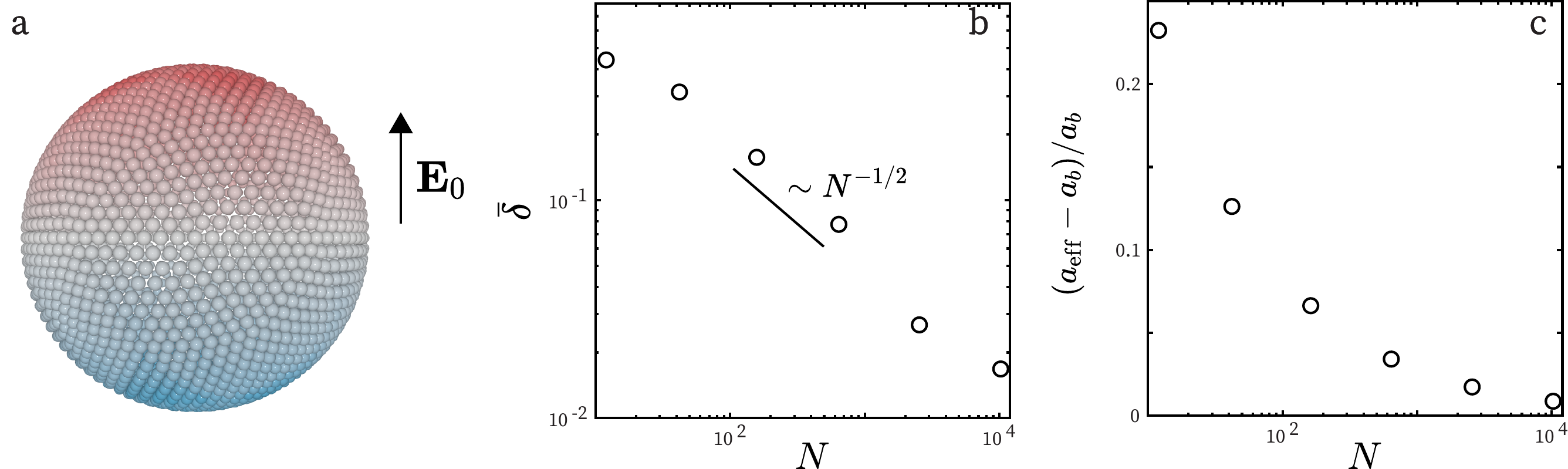}
\caption{ (\bf{a}) Snapshot of an ideally polarizable sphere comprised of $N = 2562$ beads in an electric field $\mbf{E}_0$. The induced charge distribution obtained from our immersed boundary method is indicated by the color intensity, with red and blue indicating positive and negative charges, respectively. (\bf{b}) Relative error $\bar{\delta}=\abs{\mbf{S}-\mbf{\hat{S}}}/\abs{\mbf{\hat{S}}}$ between the particle dipole $\mbf{S}$ obtained from our immersed boundary method and the exact dipole $\hat{\mbf{S}}$ in \eqref{eq:analytic_dipole} as a function of the number of constituent beads $N$. (\tbf{c}) Relative difference between the effective radius $a_\mrm{eff}$ and the true particle radius as a function of $N$.}
\label{fig:single_cond}
\end{figure*}
In this section, we demonstrate our immersed boundary method on four test problems of static configurations of polarizable colloids where exact solutions are known. 
%
%
\subsection{\label{sec:SingleConductor} Single sphere in an electric field}
A single ideally polarizable spherical particle of radius $a_{b}$ in a uniform electric field $\mbf{E}_0$ acquires a dipole moment
\begin{equation} \label{eq:analytic_dipole}
    \mbf{\hat{S}}=4\pi a_{b}^{3}\varepsilon_{\mrm{f}}\mbf{E}_0\;,
\end{equation}
In figure \ref{fig:single_cond}, we compare the dipole moment $\mathbf{S}$ computed from \eqref{eq:QS_capacitance} using our immersed boundary method on an icosphere of $N$ constituent beads to the exact $\mbf{\hat{S}}$. The relative error, $\bar{\delta}=\abs{\mbf{S}-\mbf{\hat{S}}}/\abs{\mbf{\hat{S}}}$, decays slowly as $\bar{\delta} \sim N^{-1/2}$, but an error of about $\approx 5 \%$ is obtained for $N \approx 1000$. 
One contributing factor to this error is the fact that, because tessellating beads have finite radii $a$, the total radius of the icosphere is larger than the radius of the particle sphere $a_b$ on which the bead centers lie. We define an effective radius $a_\mathrm{eff}$ of the icosphere based on the calculated induced dipole, $\mbf{S}=4\pi a_{\mrm{eff}}^{3}\varepsilon_{\mrm{f}}\mbf{E}_0$, which depends on $N$. Figure \ref{fig:single_cond}(c) shows the relative difference between $a_\mrm{eff}$ and $a_b$ for icospheres of different $N$. This difference is largest for small $N$, where the tesselating bead radii are comparable to the particle radius, and $a_\mrm{eff} \rightarrow a_b$ as the number of constituent beads increases and their radii decrease. Note that because $a_\mrm{eff}$ is always larger than $a_b$, the icosphere dipole $\mathbf{S}$ has a larger magnitude than $\abs{\hat{\mathbf{S}}}$. We prevent the errors from this single body calculation from propagating through subsequent calculations involving multiple particles by using $a_\mrm{eff}$ as the particle radius rather than $a_b$.
\subsection{Pair of spheres in an electric field}
\begin{figure*}[!t]
\centering
\includegraphics[width=0.9\textwidth]{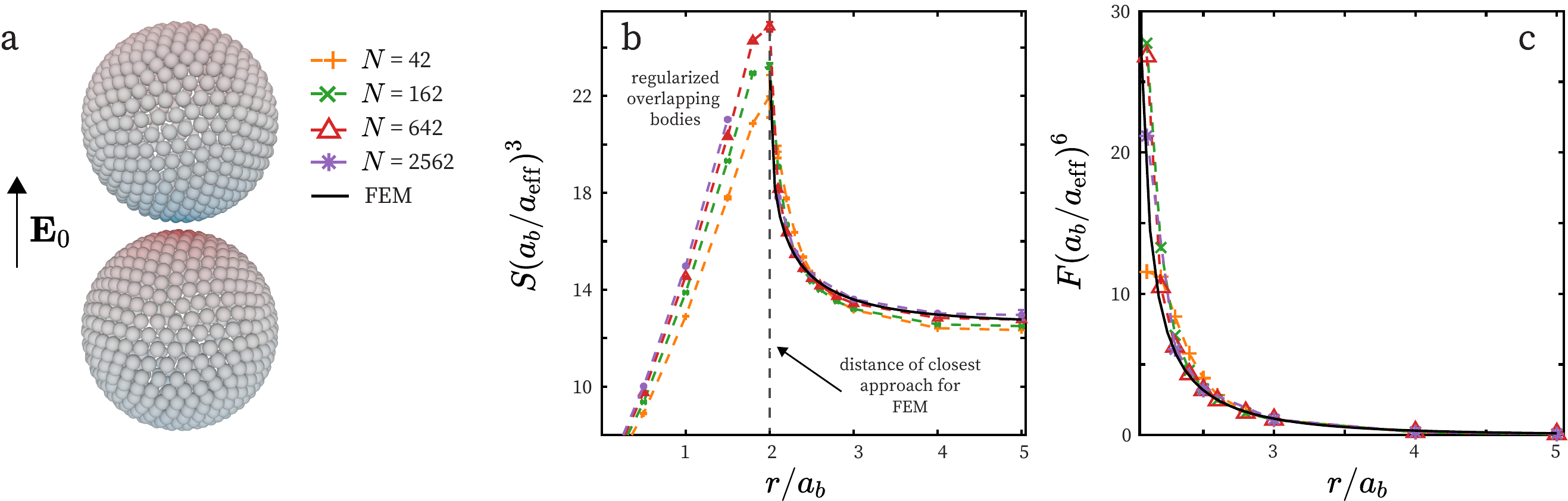}
\caption{ (\bf{a}) Pair of ideally polarizable spheres in the presence of an external field $\mbf{E}_0$. The charge distribution is qualitatively shown by color: positive and negative charges are depicted in red and blue, respectively. The saturation of the color is proportional to the absolute value of the bead's charge. ( \tbf{b} ) Computed dipole moment scaled in the effective radius of the body as a function of the center-to-center distance in units of the radius of the bodies, $a_b$. Results are shown for different levels of refinement ($N=[42,162,642,2562]$). Values obtained by the expansion in Eq. \ref{eq:analytic_twinspheresS} and recursions for the weighting coefficients in ref. \cite{Jeffrey1973} are shown as a solid black line and are equivalent to the Finite Element Method. Notice how the regularization of the charge over the surface of the rigid bodies allows for particles to overlap, that is, $r<2a_b$, without diverging. ( \tbf{c} )  Evolution of the attractive force, scaled on $a_{\mrm{eff}}$, on the spheres as the distance between them increases. The solid black line is the force obtained using eqs. \ref{eq:analytic_twinspheresS} and \ref{eq:analytic_twinspheresF}. }
\label{fig:TwinConductor}
\end{figure*}

\indent The dipoles of a pair of ideally polarizable spherical particles with radius $a_b$ in an external field $\mathbf{E}_0$ (fig~\ref{fig:TwinConductor}a) can be expressed as a power series in inverse distance\cite{Ross1968,Jeffrey1973}
\begin{equation}
    \mbf{\hat{S}} = 4\pi a_{\mrm{b}}\varepsilon_{\mrm{f}}\sum_{p=0}^{\infty} \left( \frac{a_{\mrm{b}}}{ r} \right)^{p} \left( A_{p}\mbf{I} + B_{p}\mbf{\hat{r}\hat{r}} \right)\cdot \mbf{E}_0\;,
    \label{eq:analytic_twinspheresS}
\end{equation}
where $\mathbf{r}$ is the center-to-center vector, $r = \abs{\mathbf{r}}$, $\unitvec{r}=\mathbf{r}/r$, $\mbf{I}$ is the identity tensor, and${A_p}$ and $B_p$ are coefficients found in eqs. 5.6 through 5.8 in ref. \cite{Jeffrey1973}. The force on each particle is related to the gradient of the dipole strength,
\begin{equation}
    \mbf{\hat{F}} = -\nabla \mbf{\hat{S}}\cdot\mbf{E}_0\;.  
    \label{eq:analytic_twinspheresF}
\end{equation}
\indent Figure \ref{fig:TwinConductor} shows the dipole and force on the pair of spheres computed using our immersed boundary method as a function of the distance $r$ and the number of tesselling beads $N$. The solid lines represent the analytical values computed by using eqs. \ref{eq:analytic_twinspheresS} and \ref{eq:analytic_twinspheresF}. 
Each distance point represents the average value of ten independent angular configurations for the spheres to avoid any bias introduced by the alignment of beads belonging to different bodies. However, the spread of the values at each point is small enough such that they would fall inside the markers and, therefore, is not depicted. When scaled on the effective radius, the dipole and force are in good agreement with the exact solution for a pair of spheres. \\
\indent Unlike the single-sphere case described in section \ref{sec:SingleConductor},  the charge distribution for each of the composite spheres is not symmetric about the equator. Instead, the induced charge distribution is mostly localized in the region close to the other sphere. Moreover, as the distance between surface over which the charge is distributed the dipole dramatically increases, indicating the preferred conduction path.\cite{zangwill2013} Both of these important features are shown when using the immersed boundary method with expected small deviations present for $N=42$ and only when the surface-to-surface distance is smaller than the radius of the composing bead, $a$. In the absence of hard steric repulsions, the regularization for $\psi_\alpha$ in eq. \ref{eq:pot_charge} facilitates the computation of numerically finite charge distributions of overlapping polarizable bodies. This is the case for the computed dipoles in fig. \ref{fig:TwinConductor} b for values of $r<2a_b$. In contrast, finite element-type methods are ill-equipped to estimate the induced charge of overlapping polarizable shells resulting in unphysical charge distributions. In dynamic simulations, where hard repulsions cannot be represented exactly, small overlaps are common, which makes it difficult to use finite element methods directly on configurations obtained from simulation.

\subsection{ \label{sec:ArrayConductivity} Cubic array of spheres in an electric field}
\begin{figure*}[!t]
\centering
\includegraphics[width=0.85\textwidth]{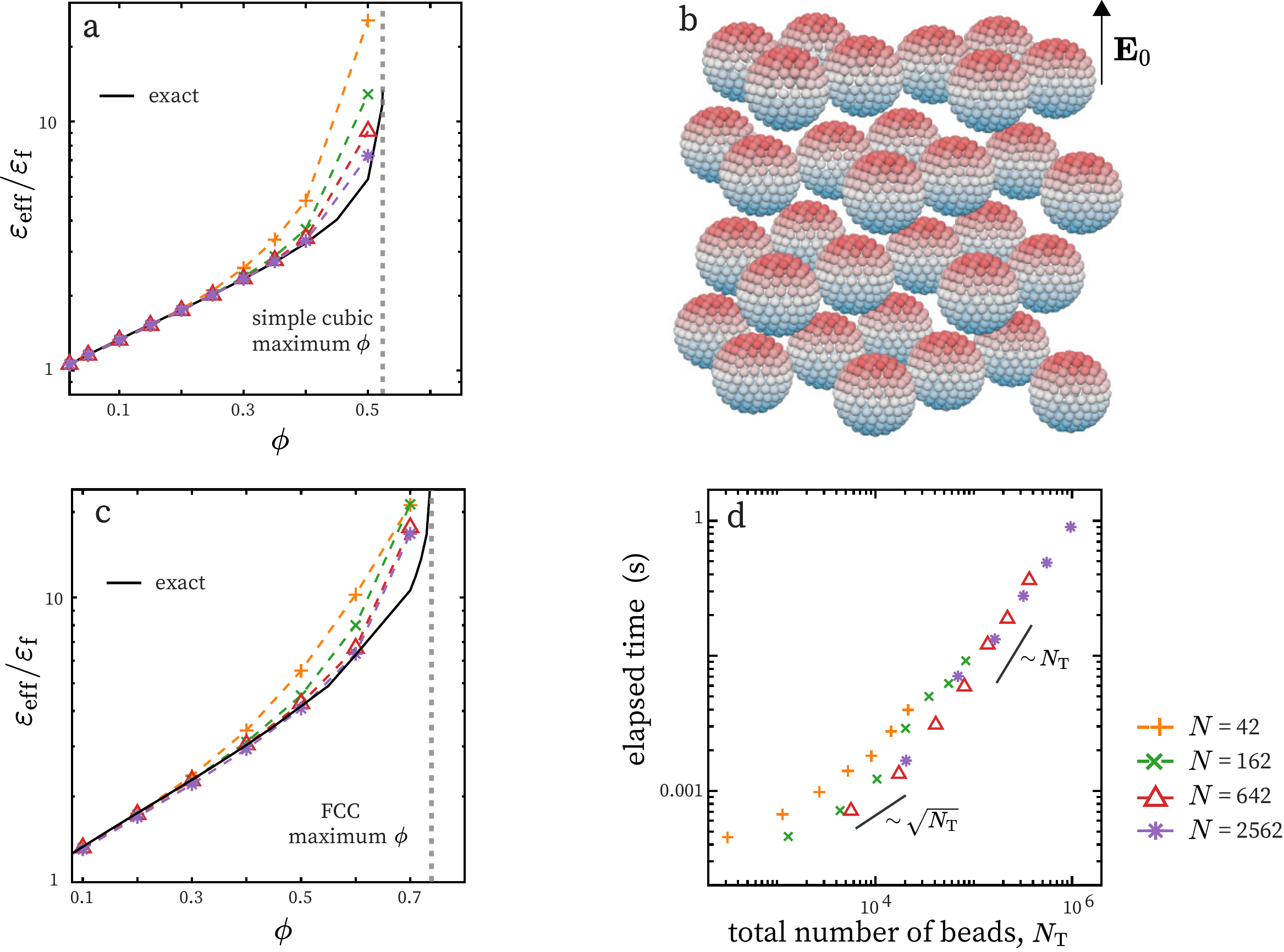}
\caption{ (\tbf{a}) Effective permittivity for a periodic simple cubic array of ideally polarizable bodies as a function of the volume fraction of the bodies. The number of beads, $N$, decreases in the downward direction ($N=[42,162,642,2562]$). The solid black line represents the results obtained by Sangani and Acrivos \cite{Sangani1983} for a simple cubic lattice.  (\tbf{b}) FCC lattice of ideally polarizable spheres in the presence of an electric field $\mathbf{E}_{0}$. The positive and negative induced charges are shown in red and blue, respectively. The saturation of the color is proportional to the magnitude of the bead charge.  (\tbf{c}) Effective permittivity for an FCC configuration of ideally polarizable bodies in the presence of an external field. The number of beads composing the bodies increases in the downward direction ($N=[42,162,642,2562]$). The black solid line shows the results obtained by 
Sangani and Acrivos for an FCC configuration of inclusions.\cite{Sangani1983}
The dashed vertical lines in (\tbf{a}) and (\tbf{c}) correspond to the simple cubic lattice packing limit, $\phi \approx 0.52$, and FCC lattice packing limit, $\phi \approx 0.74$, respectively. (\tbf{d}) Elapsed time per time step as a function of the total number of beads, $N_\mrm{T} = N_b \times N $ for FCC lattice configuration. The number of bodies in the simulation box increases to the right, $N_b = [8, 27 , 64 , 125 , 216 , 343 , 512 ]$ }
\label{fig:lattice}
\end{figure*}

To this point, we have only considered systems with a finite number of spheres. However, our goal is to take advantage of the computational method to simulate suspensions with an infinite number of particles. The simplest case is given by a periodic array of bodies. This type of two-phase media is prevalent in composite materials, and a well-studied quantity is the effective conductivity of media.\cite{Bonnecaze1990,Sangani1983,Jeffrey1973} For suspensions of ideally polarizable bodies, this conductivity is the effective permittivity of the suspension, $ \varepsilon_{\mrm{eff}} $. \cite{zangwill2013}  \\
\indent We compute the effective permittivity from the elements of the grand capacitance tensor in eq. \ref{eq:QS_capacitance} as proposed by Bonnecaze and Brady. \cite{Bonnecaze1990} That is
\begin{equation}
    \varepsilon_\mrm{eff} = \varepsilon_{\mrm{f}} - n \left\langle \mathbf{C}_{S\Psi}\cdot \mathbf{C}_{q\Psi}^{-1} \cdot \mathbf{C}_{qE} - \mathbf{C}_{SE} \right\rangle \;,
\end{equation}
where $\varepsilon_{\mrm{f}}$ is the conductivity of the particle surroundings, and $\langle \cdots \rangle$ is an average over all the bodies. The permittivity is commonly present in the tensorial form, but for an isotropic configuration, it is just a scalar value, $\varepsilon_{\mrm{eff}}\mbf{I}$. Each of the terms inside the angle brackets are the couplings between the moments of the particles and the potential and its gradients, and form the matrix 
\begin{equation}
    \mathbf{C}=\left[\begin{array}{cc}
\mathbf{C}_{q\Psi} & \mathbf{C}_{qE}\\
\mathbf{C}_{S\Psi} & \mathbf{C}_{SE}
\end{array}\right]\;.
\end{equation}
\indent In a periodic cubic array, the couplings $ \mathbf{C}_{qE}$ and $\mathbf{C}_{S\Psi}$ are zero,\cite{Bonnecaze1990}. Thus, the average dipole of the cubic array is the same as the dipole of a single sphere in a periodic cell, that is, $\varepsilon_{\mrm{eff}} = \varepsilon_{\mrm{f}}+ n\langle\mbf{C}_{SE}\rangle$. \\
\indent To assess the validity of the computed permittivities, we use Sangani and Acrivos results \cite{Sangani1983} for the effective conductivity for cubic arrays of spheres as a function of the volume fraction. \\
\indent Figure \ref{fig:lattice} (a) shows the effective permittivity in units of the fluid permittivity as a function of the volume fraction for a simple cubic array. At low concentrations, $\phi\leq 0.15$, good accuracy is obtained even for the bodies comprising a few beads. As the concentration of the bodies increases, the interparticle distance decreases and the interactions between the image bodies in the simulation cell become important. Nonetheless, good accuracy is obtained for bodies comprising $O(100)$ or more. At significantly high concentrations $\phi\geq0.40$, only bodies comprising $O(1000)$ are in close agreement with the behavior predicted by Sangani and Acrivos. Near the simple cubic maximum packing limit (vertical gray dotted line in fig. \ref{fig:lattice} (a) ) it is expected that the dipole, and consequently the effective permittivity, will diverge. Although all the composite spheres show a significant increase in the effective conductivity as we approach the packing limit, a completely accurate result would require an infinite number of moments in equation \ref{eq:Moments} which would noticeably decrease the efficiency of the method.\\
Furthermore, we computed the effective permittivity for a face-centered cubic (FCC) array from dilute to closest packing. A snapshot of one cell in the lattice is seen in figure \ref{fig:lattice} (b). The results show a similar trend to the simple cubic array, the main difference being that the closest packing for SC (indicated by the vertical dotted gray line in figs \ref{fig:lattice} (a) and (c) ) is significantly lower than for FCC.\\
\indent The GPU time complexity of our method is obtained by tracking the time required to compute the induced charge distribution for a series of FCC array configurations in an electric field using GeForce GTX 1080 graphics card by Nvidia. The level of refinement and the total number of simulated bodies varied from $N_b =  8 $ (two bodies in each dimension) to ${N_b = 512 }$ (eight bodies in each dimension). The Ewald splitting parametter was set to $\xi=0.5$ and Spectral Ewald parametter, $\eta$ chosen to commit to an error tolerance of $10^{-3}$ by relating it to the Fourier transform number of grid nodes, $P$, and spacing, $h$, via $\eta=P(h\xi)^2/\pi $.\cite{Sherman2019}  The obtained computational times are depicted in fig. \ref{fig:lattice} (d) as a function of the total number of beads in the simulation box $N_{\mrm{T}}=N \times N_b$. The distribution of elapsed times shows a time complexity $O(\sqrt{N_{\mrm{T}}}\log N_{\mrm{T}})$ for simulations containing up to $O(10^5)$ beads. Although log linearity is reached for larger systems, the time required to obtain the charge distribution of simulations containing up to ${O(10^{6})}$ beads remains within one second.
\subsection{ \label{sec:RigidRod} Rigid rod in an electric field}
\begin{figure*}[!t]
\centering
\includegraphics[width=0.9\textwidth]{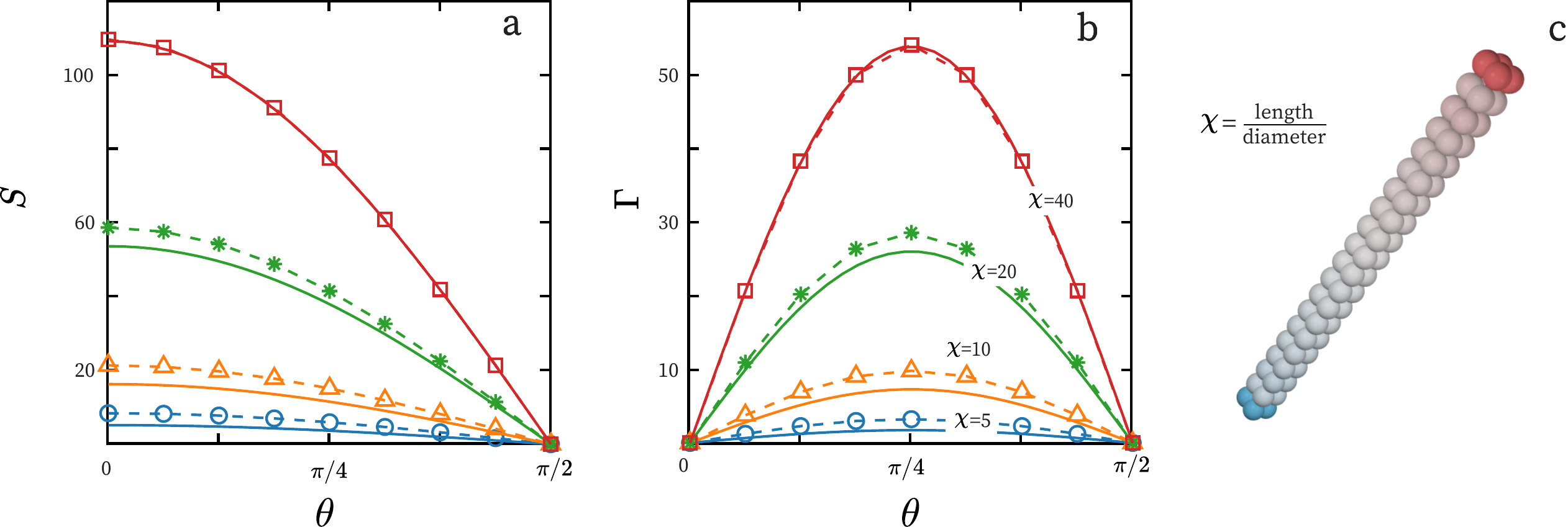}
\caption{ Evolution of the ( \bf{a} ) induced dipole moment in the direction of the external electric field and ( \bf{b} ) resulting torque as a function of the angle between the applied field and the longest axis of the fiber at different values of aspect ratio, $\chi$. The aspect ratio increases in the upwards direction. Values corresponding to $\chi=5$, $\chi=10$, $\chi=20$ and $\chi=40$ are represented by circled (in blue), triangled (in orange), starred (in green), and squared (in red) markers, respectively. The solid lines in panel (\bf{a}) and (\bf{b}) are computed using the depolarization tensor, as in eq. \ref{eq:depolarization}, for a prolate spheroid. \cite{torquato_random_2002} }
\label{fig:FiberConduct}
\end{figure*}
Colloidal particles are present in a variety of topologies. Our immersed boundary method is not only applicable to spherical bodies, but can include other geometries as well. Consider, for example, a long but slender ideally polarizable body, hereafter simply a ``fiber'', immersed in an external electric field.  \\
\indent The fiber's geometry can be approximated by a variety of structures. For the purpose of this discussion, we approximate the body by four parallel lines of particles composing a single fiber. The anisotropy of the fiber is controlled by the ratio of the longest to the shortest axis, $\chi$. Figure  \ref{fig:FiberConduct} ( c ) shows this structure in the presence of an external field.\\
While we are not aware of the existence of an exact solution for this particular geometry, we expect the dipole moment and torque on the fiber to converge to that of a prolate spheroid for large values of $\chi$. The dipole moment of an ideally polarizable body can be found via the relation \cite{torquato_random_2002} 
\begin{equation}
 \mbf{S} = \left[ \mbf{I} - \mbf{A}^\star \right]^{-1} \cdot \mathbf{E}_0;\quad \quad   \mathbf{A}^{\star}=\left[\begin{array}{ccc} Q & 0 & 0\\ 0 & Q & 0\\ 0 & 0 & 1-2Q \end{array}\right]\;.
    \label{eq:depolarization}
\end{equation}
Where $\mathbf{A}^{\star}$ is known as the depolarization tensor, \cite{Jackson1999} and $Q$ for a prolate spheroid is given by
\begin{equation}
Q=\frac{1}{2}\left\{ 1+\frac{1}{(b/a)^{2}-1}\left[1-\frac{1}{2\chi}\log\left(\frac{1+\chi}{1-\chi}\right)\right]\right\}\; .
\end{equation}
\indent Figure  \ref{fig:FiberConduct} ( a ) and ( b ) show the induced dipole moment on the fiber, $S$, and the resulting torque, $\Gamma=\mbf{S}\times \mbf{E}_0$ as a function of the angle between the applied field and the longest axis of the fiber. The ratio of the longest to the smallest axis, $\chi$, ranges from $\chi=5$ to $\chi=40$. Solid lines represent the computed dipole moments and torques using eq. \ref{eq:depolarization}. Notice that the induced dipole moment in the direction of the field is at its maximum when the field and the longest axis are parallel to each other ( $\theta=0$ ) and decreases to zero when they are orthogonal to each other ( $\theta=\pi/2=90^o$ ). Furthermore, the maximum of the torque, fig.  \ref{fig:FiberConduct} ( b ) is found at values of $\theta=\pi/4=45^{o}$ as expected for this geometry. At low values of $\chi$, significant differences are obtained between the dipole moment computed by the simulations and the dipole moment obtained by the polarization tensor. This is a result of the geometrical dissimilarities between our approximated fiber and the prolate spheroid. Nonetheless, as the ratio, $\chi$, increases in value, the far-field of both structures tends to the same dipole and torque, which corresponds to the limit where these dissimilarities are small.
\section{\label{sec:BDofColloids} Brownian dynamics simulations of charged and polarizable colloids in explicit electrolytes}
The numerical computations presented in \ref{sec:StaticSims} provide information on the numerical accuracy of our immersed boundary method when computing single- and multibody properties of equally sized bodies. However, the primary goal of our simulation method is to accurately and efficiently simulate electrostatically stabilized colloidal suspensions. In this section, we will study the method in the context of Brownian Dynamics simulations of colloidal particles in the presence of an explicit electrolyte. First, we show the equilibrium structural properties of the electric double layer for a negatively charged colloidal particle in an explicit electrolyte. Then we demonstrate the applicability of the method to obtain transport properties of colloidal suspensions by computing the electrophoretic velocity of charged colloids and studying the electrokinetic phenomenon known as ``induced-charge electroosmosis''.\cite{Bazant2004}

\subsection{\label{sec:BD_methods} Brownian dynamics}
\indent We model the ions and counterions in the electrolyte as a suspension of charged, hard, spherical particles. We will interchangeably refer to them as ``beads.'' Colloidal particles are constructed as explained in Section \ref{sec:IMBOUND} and we will commonly call them ``bodies.''  The solvent is treated implicitly and is assumed to be Newtonian, so that it interacts with the ions through hydrodynamic forces from flows in the medium and stochastic Brownian forces due to momentum relaxation of the solvent molecules. \cite{Frenkel2002} On the colloidal scale, inertial relaxation occurs on time scales orders of magnitude smaller than those on which the ions move. In this regime, any perturbation to ion momentum is felt almost instantaneously, which causes the ions to move at its terminal velocity. Under these assumptions, inertial effects can be ignored and the overdamped Langevin equation governs the dynamics of the ions:\cite{Frenkel2002}
\begin{equation}
0=\mathbf{F}_{\alpha}^{\mathrm{H}}+\mathbf{F}_{\alpha}^{\mathrm{I}}+\mathbf{F}_{\alpha}^{\mathrm{E}}+\mathbf{F}_{\alpha}^{\mathrm{B}}\:,
\label{eq:overdamped}
\end{equation}
where $\mathbf{F}_{\alpha}^{\mathrm{H}}$ is the hydrodynamic force acting on the
$\alpha^{\mathrm{th}}$ ion, $\mathbf{F}_{\alpha}^{\mathrm{I}}$ accounts for forces arising from a generic conservative potential, $\mathbf{F}_{\alpha}^{\mathrm{E}}$ is the external force exerted by a global electric field \cite{Sherman2019} and $\mathbf{F}_{\alpha}^{\mathrm{B}}$ is the stochastic Brownian force. The last force satisfies the fluctuation-dissipation theorem \cite{Russel1989} with ensemble average:
\begin{align}
    \left\langle \mathbf{F}^{\mathrm{B}}(t)\right\rangle &=0; \\ \nonumber
    \left\langle \mathbf{F}^{\mathrm{B}}(t)\mathbf{F}^{\mathrm{B}}(t+\tau)\right\rangle &=2k_{\mathrm{B}}T(\mathbf{M}^{\mathrm{H}})^{-1}\delta(\tau),
\end{align}
where $ \mathbf{F}^B(t) = [ \mathbf{F}^{\mathrm{B}}_{1}(t), \mathbf{F}^{\mathrm{B}}_2(t), \ldots ] $, $ \left\langle \cdot\right\rangle $ indicates the expectation value, $\delta$ is the Dirac delta function, and $\mathbf{M}^{\mathrm{H}}$ is the hydrodynamic mobility tensor. This formulation ensures that any energy an ion gains from a thermal fluctuation is dissipated as drag to the solvent.\\
\indent The hydrodynamic mobility tensor couples the non-hydrodynamic force, $\mathbf{F}_{\beta}=\mathbf{F}_{\beta}^{\mathrm{I}}+\mathbf{F}_{\beta}^{\mathrm{E}}+\mathbf{F}_{\beta}^{\mathrm{B}}$, to the velocity of the $\alpha^{\mathrm{th}}$ ion:
\begin{equation}
   \mathbf{u}_{\alpha}(t)= \sum_{\beta=1}^{N}\mathbf{M}_{\alpha \beta}^{\mathrm{H}}\cdot\mathbf{F}_{\beta}(t)\;.
   \label{eq:velocity}
\end{equation}
\indent For simplicity, we will neglect interparticle hydrodynamic interactions, but these can be included using the Rotne-Prager-Yamakawa mobility tensor.\cite{SwanGang2016} Thus, the drag on each ion is decoupled from all the others and is equal to the Stokes drag,  
\begin{equation}
    \mathbf{M}_{\alpha \beta}^{\mathrm{H}}=0,\alpha \neq \beta ;\:\hspace{4mm}\mathbf{M}_{\alpha \alpha}^{\mathrm{H}}=\mathbf{I}/\gamma\;,
\end{equation}
where all ions are assigned the same drag coefficient, $ \gamma $.  Equation \ref{eq:overdamped} can be solved numerically via a Euler-Maruyama integration scheme:
\begin{equation}
\mathbf{x}_{\alpha}(t+\Delta t)=\mathbf{x}_{\alpha}(t)+ \mathbf{u}_{\alpha}(t)\Delta t\:,
\label{eq:forward}
\end{equation}
where $ \Delta t $ is the time step over which the ion trajectories advance.\\
 \indent Forces arising from conservative interactions among ions are represented as the gradient of a potential energy $U(\mathcal{X})$, which is a function of the coordinates of all ions $\mathcal{X} \equiv [\mathbf{x}_1, \mathbf{x}_2, \dots, \mathbf{x}_N]^T$,
\begin{equation}
\mathbf{F}_{\alpha}^{\mathrm{I/E}}(\mathcal{X}) \equiv -\nabla_{\mathbf{x}_{\alpha}} U^{\mathrm{I/E}}(\mathcal{X}),
\label{eq:potential_grad}
\end{equation}
where the gradient is taken with respect to the position of the $\alpha$\textsuperscript{th} particle.  \\
\indent In our analysis, we are concerned with finite sized ions, which cannot overlap. The hard-sphere force computed by the derivative of the well-known step-like potential is discontinuous; it is zero everywhere except for a $\delta$-function of infinite magnitude at contact. This type of potential cannot be implemented directly in simulations.  Typically, the hard potential is approximated with a soft potential of the form $r^{-n}$, where $n$ is a large power.  The larger $n$ is, the more accurately the soft potential approximates the hard potential, but the resulting force becomes larger as the potential diverges increasingly rapidly. \cite{Heyes1994A}  Smaller time steps must therefore be taken as $n$ increases to prevent unphysically large steric forces, rendering this method computationally inefficient.  Heyes and Melrose implemented a "potential-free" hard sphere algorithm by allowing particles to overlap over the course of a time step because of other forces and then separating them to contact at the end of the time step. \cite{Heyes1993}  Because equations \eqref{eq:velocity} and \eqref{eq:forward} give a relation between particle displacements and forces, we can compute the effective force required to move two overlapping ions back into contact after one time step. Thus, the potential-free algorithm can be equivalently written in terms of a hard sphere pair potential:\cite{Varga2015}
\begin{equation} \label{eq:hardsphereFD}
U_{\alpha \beta}^{\mathrm{hs}}(r)=\begin{cases}
\frac{\gamma}{4\Delta t}(r-2a)^{2} & \text{if }r<2a\\
0 & \text{if }r\geq2a
\end{cases} \: ,
\end{equation}
where $a$ is the hard-sphere radius of an ion. In the form of the potential used here, ${\gamma}$ and ${a}$ are the same for all particles.\\ 
%
\indent All simulations are of colloids immersed in the binary and symmetric electrolyte; that is, ions of different species have the same charge in magnitude but opposite in sign. Distance and time are made dimensionless by measuring distances relative to the radius of the ion, $a$, and time relative to the diffusion time of the ion, $\tau_{\mrm{D}}=k_{\mrm{B}} T /\gamma a^2 $. The charge scale is set by $\sqrt{\varepsilon_{\mathrm{f}} a k_{\mathrm{B}} T} $. The volume fraction of the colloids, $\phi_b$, and the length of the simulation box are the same in all simulations and given by $\phi_b =0.01$ and $L=108a$, respectively. The charge of the ions, $q_i$, is prescribed by the dimensionless parameter $\epsilon$ (different from the permittivity of the solvent, $\varepsilon_\mrm{f}$). We define this quantity as the electric potential between two charges at contact with respect to the thermal energy, 
\begin{equation}
   \epsilon=\frac{q_{i}^2}{8 \pi \varepsilon_{\mathrm{f}}a k_\mathrm{B} T}  \;.
\end{equation} 
The presented values of the electrostatic interactions relative to ${ k_{\mathrm{B}}T }$ are chosen to be within the range of values commonly found in common electrolytes: ${\epsilon = [0.5, 2.0 ] }$. For example, for a solution of ${\mathrm{NaCl}}$ in water at room temperature, ${\epsilon\approx1.1}$ (assuming relative permittivity of $80$). The ratio of the charge in the colloid with respect to the ions is given by $\sigma= Q/(N q_i)$, where $Q$ is the total charge in the colloid and $N$ the number of beads that compose the body. The charge in the colloid needs be balanced by $\sigma N$ counter-ions to ensure electroneutrality. Simulations were at least $200\tau_{\mrm{D}}$ long and the first $100\tau_{\mbf{D}}$ are discarded to ensure equilibrium. The integration time-step was $10^{-3}\tau_{\mrm{D}}$ and of equal value in all simulations. 
\subsection{\label{sec:equilibrium_simulations} Electric double layer around a charged colloid}
\begin{figure*}[!t]
\centering
\includegraphics[width=0.9\textwidth]{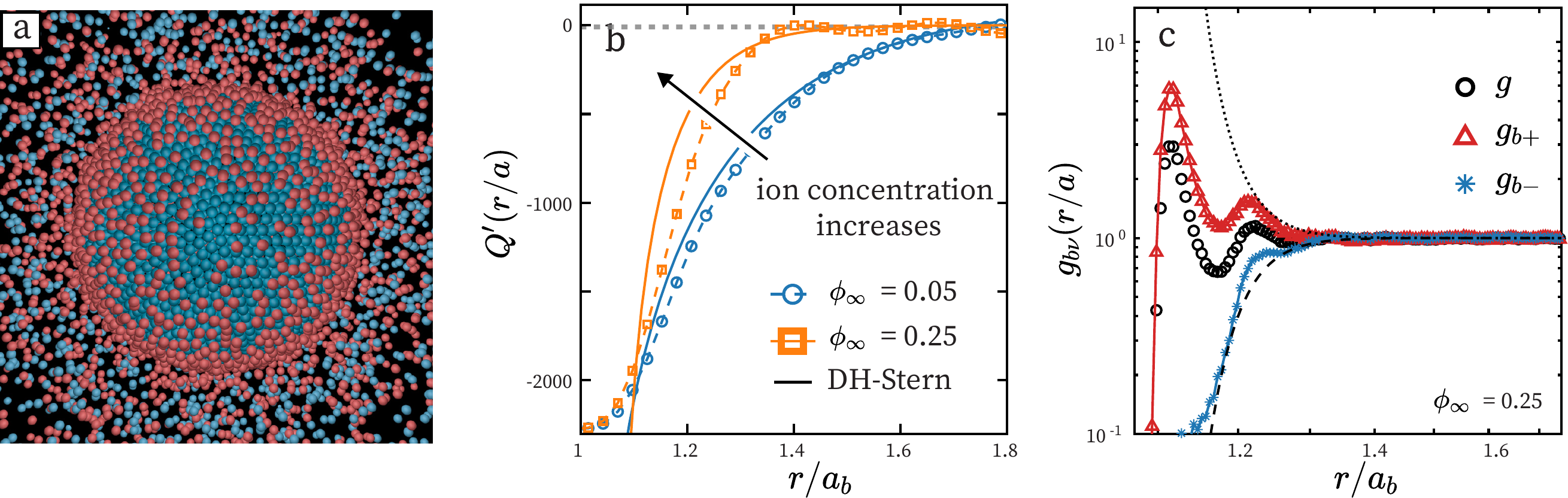}
\caption{ (\bf{a}) Snapshot of charged colloid in an electrolyte depicting counterions being attracted to the surface of the colloidal particle (some ions were removed for clarity). Positively charged and negatively charged species are shown in red and blue. (b) Cumulative charge density, as present in eq. \ref{eq:Qtot_sim}, from dynamic simulations of negatively charged colloid in an explicit electrolyte at different ion volume fractions. Circles in blue and squares in yellow represent concentrations of $\phi = 0.05$ and $\phi = 0.25$, respectively. Solid lines represent the expected cumulative charge distributions computed using Debye-Huckel theory and considering the Stern layer; the Stern layer thickness was left as a fitting parameter. (c) Colloid-ion correlation functions (black markers), colloid-counterion (solid red line with triangular markers), and colloid-co-ion pair correlation functions at different ion volume fractions showing the microstructure of the double layer at $\phi=0.25$. Dotted and dashed lines show the expected colloid-counterion and colloid-coion pair correlation functions, respectively. The distributions are computed by assuming that the pair correlation function can be approximated by $g_{i j } \sim \exp\left( -U_{i j }(r)\right)$. }
\label{fig:SnapshotGofR}
\end{figure*}
\indent Our immersed boundary method can be used to identify important features of electrostatically stabilized colloidal suspensions. We illustrate some of our method's capabilities by studying the structure of the electric double layer present on the colloid-solvent interface. Figure \ref{fig:SnapshotGofR} (a) shows a snapshot of a negatively charged colloidal particle in an explicit electrolyte, where some of the ions are not shown, allowing qualitative details of the counterion cloud to be easily observed. The counterions generally try to balance the charged colloid and acquire electroneutrality. Using the positions of the ions, we can gain some insight into the charge density distribution, $\rho(\mbf{r})$, and consequently verify the electroneutrality of the simulation box. Figure \ref{fig:SnapshotGofR} (b) shows the cumulative charge density,
\begin{equation}
Q^{\prime}(r)=Q+\int_{a_{b}}^{r}\left\langle \rho(r^{\prime})\right\rangle \mathrm{d}r^{\prime}\;,
\label{eq:Qtot_sim}
\end{equation}
at different ion volume fractions, $\phi_\infty = [0.05,0.25]$. The strength of the ion-ion interactions at contact for the simulations shown was set to $\epsilon=0.5$ in units of the thermal energy, $\kb T $. The colloid is made up of 642 beads, has a radius of $a_{b}\approx14.5a$ and $\sigma=1.0$. Solid lines show the expected behavior of the Debye-Huckel-Stern theory,\cite{Russel1989} where the Stern layer thickness was left as a fitting parameter.

The local charge densities required to compute $Q^{\prime}(r)$ are found by discretizing the space surrounding the colloid into consecutive shells. The average local charge at each of the discretization points,$\left\langle \rho(r)\right\rangle$, was found using a sliding window algorithm.\cite{proakis1996} We chose the window to be twice the radius of the ion and the discretization steps equal to $0.1a$. \\
\indent Consistent with the screening of electrostatic interactions, the decay of the cumulative charge is proportional to the volume fraction of the ions ($\phi_\infty = [0.05,0.25]$). Notice significant differences between the computed cumulative charge using Debye-Huckel-Stern theory, and that computed from the simulations, particularly at high concentrations where screening effects give rise to long-range charge density oscillations \cite{ekrucker2020}.\\
\indent While the cumulative charge and other integrated quantities (such as capacitance) can be well estimated using local density approximations to obtain the charge density distribution of the ionic species near charged surfaces,\cite{ Squires2015, Khair2017} to our knowledge, these cannot correctly represent structural features on the order of the ion diameter of the double layer structure. The local microscopic density or structure of a liquid (and other states of matter in other contexts) is better characterized by the partial pair correlation function, which for species $\nu$ and $\mu$ is defined as:\cite{hansen_mcdonald_2006}
\begin{equation}
    g_{i j }(\mathbf{r})=\frac{1}{Vn_{i}n_{j}}\left\langle \sum_{\alpha=1}^{N_{i}}\sum_{\beta \neq \alpha}^{N_{j}}\delta(\mathbf{r}+\mathbf{r}_{\alpha}-\mathbf{r}_{\beta})\right\rangle\:,
    \label{eq:binary_gofr}
\end{equation}
where ${\mathbf{r}_{\alpha/\beta}}$ is the position of particle ${\alpha}$ of species ${i}$ or of particle ${\beta}$ of species ${j}$.  In single-component systems, particles $\alpha$ and $\beta$ belong by definition to the same species. For multicomponent systems, the species $i$ and $j$ can represent the same or different species.\\
\indent Figure \ref{fig:SnapshotGofR} ( c ) presents an example of the pair correlation functions between 1) the colloid and all ionic species, $g(r)$, 2) colloid and counterions, $g_{b+}(r)$, and 3) colloid and coions, $g_{b-}(r)$ for volume fractions of $\phi_\infty = 0.25 $. The pair correlation function is obtained using the standard ``shell summation'' method\cite{kjellander_2001} averaged  over 50 uncorrelated configurations over time. The simulation's results are compared against the values obtained by approximating the pair distribution function to be $g_{i j }\sim\exp(-U_{i j }(r))$. \cite{goodstein2014} The interparticle potential, $U_{i j }$, is a screened Coulomb potential between two small charged molecules. \cite{israelachvili2011} This approximation for the partial pair correlation function is only valid in the low concentration regime and in the limit of $ r \gg a_{b} $, so deviations near the surface of the colloidal particle are expected. The first peaks of $g$ and $g_{b,+}$ (shown in red and black markers), are approximately located at the distance of closest approach between the ions and the colloidal particle ( $r\approx 16.5a$ or $r\approx 1.1a_b$ ). A second shell of counterions (located at $r\approx 18.5a$ or $r\approx 1.2a_b$) surrounding the colloid is also present. The colloid-co-ion pair correlation function shows a depletion of like-charged ions near the surface of the colloid. Note that $g_{b-}$ is nonzero for values of  $r\approx 1.15a_b$; as screening effects and excluded volume interactions become important, the repulsive force between the colloid and the like charged ion is opposed by additional osmotic forces from the surrounding ions, likely giving rise to this nonzero value of $g_{b-}$.
\subsection{\label{sec:Electrophoresis} Induced-charged electrophoresis}
\begin{figure*}[!t]
\centering
\includegraphics[width=0.85\textwidth]{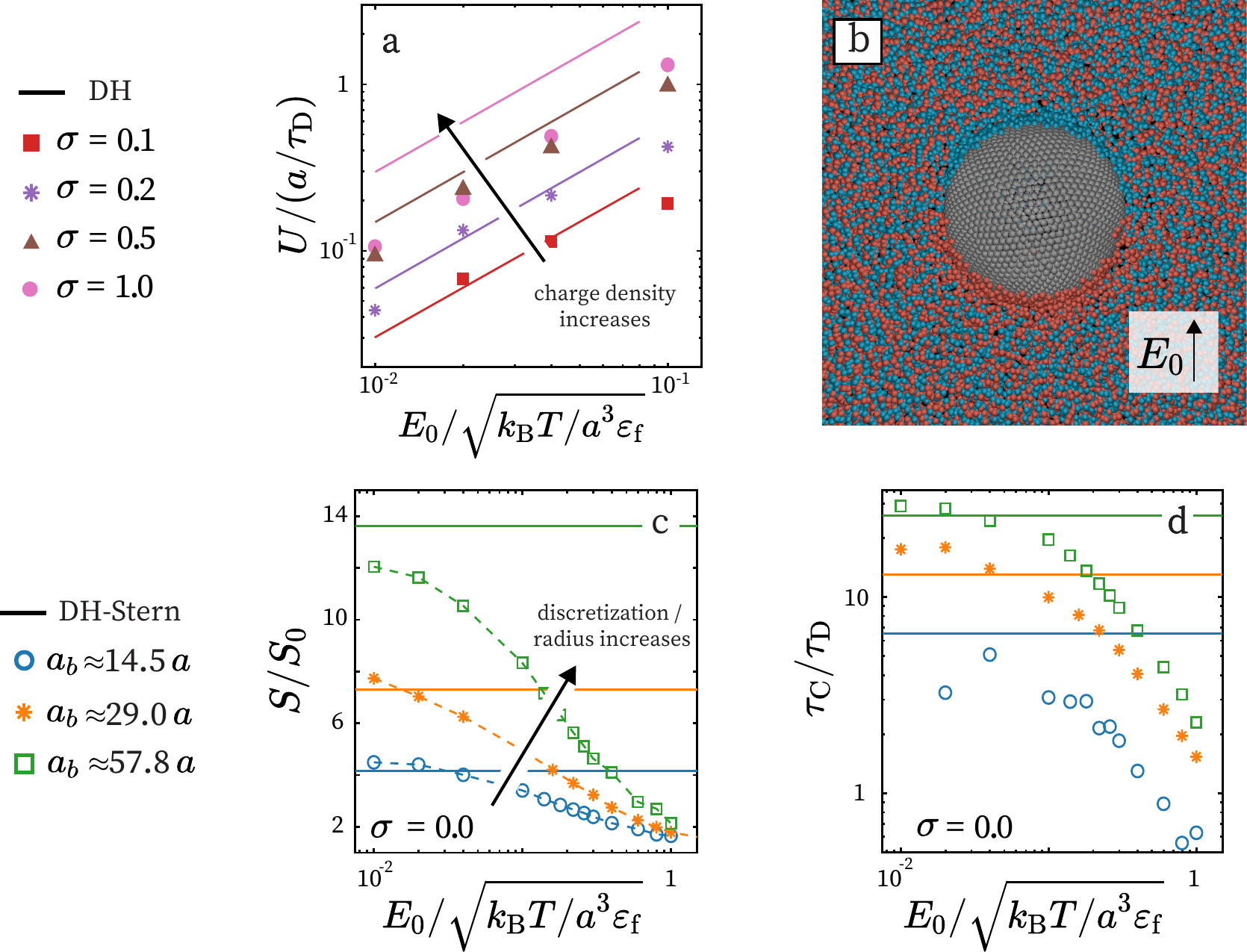}
\caption{ (a) Snapshot showing the induced charge and the resulting neutralizing double layer for an ideally polarizable body in an explicit electrolyte. The resulting double-layer structure is formed by electrosmotic effects due to the induced dipole moment of the ideally polarizable colloid. ( b ) Electrophoretic velocity as a function of the electric field. The results of charge densities $q_0 = 0.12$ (square markers), $q_0 = 0.24$ (starred markers), $q_0 = 0.62$ (triangle markers), and $q_0 = 1.22 $ (circled markers), are shown in red, purple, brown, and pink, respectively. The solid lines correspond to the computed velocities in the thin double layer limit using eq. \ref{eq:phor_vel}  (c) Evolution of the Body's induced dipole moment and (d) charging time of the electric double layer as a function of the externally applied field. The circled, starred, and squared markers correspond to discretization levels of $a_{b} = 15 a $, $a_{b} = 30 a $, and $a_{b} = 60 a $ (comprising 42, 162, and 642 beads), respectively. Solid lines are obtained using the Debye-Hückel-Stern approximation.}
\label{fig:ICEOsnap}
\end{figure*}

When the colloidal particle has a non-zero net surface charge, the presence of an external electric field will cause the colloidal particle to translate in the electrolyte with a velocity,
\begin{equation}
    \mbf{U} = \frac{\varepsilon_\mrm{f} \zeta_0}{\mu} \mbf{E}_0\; ,
    \label{eq:phor_vel}
\end{equation}
where $\zeta_0$ is the net zeta potential and $\mu$ is the solvent viscosity. Eq. \ref{eq:phor_vel} holds if the electric double layer thickness is significantly smaller than the sphere radius; that is, the thin double layer limit. Figure \ref{fig:ICEOsnap} (a) shows the electrophoretic velocity for a colloidal particle in an electrolyte for different charge densities. The filling fraction of the electrolyte is $\phi_\infty=0.10$ and the radius of the rigid body, $a_b$, is approximately $29a$ (or $N=642$ composing beads). As expected, the velocity obtained in the simulations increases with the applied electric field and the charge density. Importantly, the predicted electrophoretic velocity from eq. \ref{eq:phor_vel}, depicted by the solid lines, is in close agreement with the simulations for the lowest charge densities (starred and squared markers), which is the limit where this equation is valid. The highest charge density values (triangled and circled markers) show greater deviations from the ideal solution velocity, likely due to particle and double layer convection.
%
%
\indent In the presence of an applied electric field, polarizable colloidal particles acquire a non-uniform surface charge distribution in addition to its fixed charge. When surrounded by ions in solution, positive ions are drawn toward the particle's negative induced charges, while negative ions are drawn toward the positive induced charges on the surface of the colloid. That is, the ion cloud of the colloid also polarizes, as seen qualitatively in the ion cloud distribution in Fig. \ref{fig:ICEOsnap} (b) . The overabundance of ionic species near the surface of the colloid creates an \emph{additional} potential gradient. This ion-dense fluid contributes further to the polarization of the colloid. During this process, the colloidal particles and the double layer region ``charge up'' over time. The charge distribution emanates a quadrupolar flow field in the surrounding fluid. The charging dynamics can be explained using Squire and Bazant's electrokinetic theory of ``induced-charge electroosmosis'' (ICEO) phenomenon. \cite{Bazant2004, squires2004} Assuming ideal solution behavior, the charging time and dipole moment are found to be
\begin{equation}
    \tau_\mrm{c} = \frac{\varepsilon_\mrm{f} \kappa_{\mrm{D}} a_{b} }{2 \sigma_i (1+\kappa_{\mrm{D}} a)}; \quad  \quad \frac{S}{S_0} = 1 + \frac{ \kappa_{\mrm{D}}a_b  }{2 \left( 1+\kappa_{\mrm{D}}a \right) }\; ;
    \label{eq:ICEO_DH}
\end{equation}
where $\sigma_i$ is the ionic conductivity, different from the ratio of the colloid bead charge to the ion charge, $\sigma$. \\
 \indent This electrokinetic phenomenon behaves differently from classical electrophoresis due to the presence of the polarized double layer, which negatively aligns with the external field. In fact, this polarizable double layer becomes unstable above a certain critical electric field. Small thermal fluctuations can break the symmetry of the polarized ion cloud and result in a spontaneous rotation of colloidal particles in the orthogonal direction of the applied field.\cite{khair_lift_2019,sherman_spontaneous_2020} This spontaneous rotation is known as the Quincke rotation\cite{jones1984}. When coupled with the motion of induced charge electrophoresis, this phenomenon is analogous to the ``Magnus effect'' at larger Reynolds numbers. Here, we limit our analysis to applied electric fields below the instability threshold, but the Immersed Boundary Method described in this document has previously been used to study the ``electrokinetic magnitude effect''. \cite{sherman_spontaneous_2020}\\
\indent We performed simulations at various discretization levels and followed the dependence of the dipole moment of the particle as a function of the magnitude of the applied field.  Figures \ref{fig:ICEOsnap} (\tbf{c}) and \ref{fig:ICEOsnap} (\tbf{d}) show the final dipole strength and charging time, $\tau_{\mrm{c}}$, in units of $\tau_{\mrm{D}}$ as a function of the applied field scaled on the electric field scale set by $\varepsilon_{\mrm{f}}$, $\sqrt{k_{\mrm{B}}T/a^3\varepsilon_{\mrm{f}}}$. The blue circled, yellow starred and green squared markers represent the values obtained from colloidal particles of radius $a_b \approx 14.5a$ (composed of $N=642$ beads), $a_b \approx 29a$ (composed of $N=2562$ beads) and $a_b \approx 57.8a$ (composed of $N=10,242$ beads), respectively.  Solid lines correspond to the dipole moment and charging time computed using eq. \ref{eq:ICEO_DH} and Debye-Huckel-Onsager relation for the ionic conductivity conductivity.\cite{AvniAndelman2022} 
It can be seen in the plots that the simulations and the ICEO model agree qualitatively. Figure \ref{fig:ICEOsnap} shows the particle dipole and the charging time for particles with various net charges from simulation and Carnahan-Starling theory. As the differential capacitance decreases with $\zeta$ and $q$ grows slowly,\cite{Sherman2019}  particles with a net charge begin with a lower capacitance than a charge-free particle and charge less strongly.  Both the dipole strength and the charging time decrease with $\sigma$ at small $\widetilde{E}_0=E_0/\sqrt{k_{\mathrm{B}}T/a^3\varepsilon_{\mathrm{f}}}$.  In fact, charging is nearly completely suppressed for the largest $\sigma$ tested.  The net charge is important when its associated zeta potential is comparable to or greater in value than the induced zeta potential, $\zeta_0 \gtrsim a E_0 $.  As $E_0 $ increases, the induced $\zeta$ dominates the net $\zeta_0$ and all the net charged particles behave the same.
\section{Conclusion}
\indent We have presented an immersed boundary method for simulating equilibrium and dynamic properties of charged and/or polarizable colloidal dispersions. Our method utilizes a Spectral Ewald summation and is parallelized for computation on GPUs, allowing for robust error control while efficiently scaling to $O(10^6)$ beads. We have shown that our method is a useful tool for modeling a variety of soft electokinetic materials, including collections of polarizable colloidal particles (Sec.~\ref{sec:SingleConductor} -\ref{sec:ArrayConductivity}) and anisotropic (Sec.~\ref{sec:RigidRod}) colloidal particles in electric fields, ionic double layers around charged colloids in electrolytes (Sec.~ \ref{sec:equilibrium_simulations}), and ion and colloid dynamics of electrolytic dispersions under an applied voltage (Sec.\ref{sec:Electrophoresis}).\\
\indent We have computed the pair correlation functions and charge density distribution of the electric double layer in which our simulations show similar behavior to mean-field treatments of the electrolyte in the region where ideality assumptions are valid (low concentration, weak electrostatic interactions). Importantly, we find qualitative and quantitative improvements over those obtained from mean-field simulations for concentrated and highly correlated systems where long-range ordering of the charge density distribution arises. The method is able to accurately estimate transport properties of electrostatically stabilized colloidal dispersions and accurately represent nonlinear electrokinetic phenomena. To the best of our knowledge, existing particle-based methods look only at the equilibrium fluctuations via Green-Kubo relations to obtain transport properties of the charged species. \cite{mcquarrie2000, shi_self-diffusion_2005} Nonetheless, the transport coefficients obtained from simulations near equilibrium cannot be used to predict the response to alternating electric fields, nor do they consider any induced-charge polarization effects (such as induced charge electrophoresis).\cite{Bazant2004} \\ 
Finally, an existing GPU-capable molecular dynamics package, HOOMD,\cite{anderson_general_2008} was leveraged for thermodynamic modeling of rigid assemblies and data handling. A plug-in for HOOMD can be found in the supporting material distributed under the GPL license, which can be used to validate all of the shown results. Moreover, the plug-in can be used to provide a reliable picture of structural and transport properties of particles of arbitrary shape, such as rigid rods and rings. \\
Overall, our spectrally accurate GPU-driven immersed boundary method for simulations of ideally polarizable bodies is optimal for particle-based simulations of charged soft matter system with simulation boxes comprising anywhere from a few thousand to low millions beads.
\section*{Supporting Material}
The data that support the findings of this study are available from the corresponding author upon reasonable request. The plug-in can be downloaded in \href{GitHub}{https://github.com/Swan-Group-Code/ConductorRigid} by accessing the following URL: \url{https://github.com/Swan-Group-Code/ConductorRigid}

\section*{Acknowledgments}
We wish to thank Alfredo Alexander-Katz, Madhu Majii, and Martin Z. Bazant for their helpful discussions.
We wish to acknowledge the support from NASA, Grant No. 80NSSC18K0162 and NSF, Career Award No. 1554398. 

\section*{Appendix}
\subsection*{ Summation Tensors}
The summation tensor, $\boldsymbol{\Sigma}$, in eq. \ref{eq:saddle} is an $N_b\times N$ matrix whose rows correspond to rigid bodies and columns correspond to beads:\\ 
\begin{equation} %
\label{eq:Sigma1}
\boldsymbol{\Sigma} \equiv 
\begin{bmatrix}
\bovermat{ $N_{1}$ }{1 & 1 & \cdots} & \bovermat{$N_{2}$}{0 & 0 & \cdots} & \bovermat{$N_{b}$}{0 & 0 & \cdots} \\
0 & 0 & \cdots & 1 & 1 & \cdots & 0 & 0 & \cdots \\
0 & 0 & \cdots & 0 & 0 & \cdots & 1 & 1 & \cdots \\
\vdots & \vdots &  & \vdots & \vdots &  & \vdots & \vdots & 
\end{bmatrix}\;,
\end{equation}
 Each row is entirely comprised by $0$s, except for the $N_i$ consecutive $1$s corresponding to the $N_i$ beads in the rigid body $i$.  \\
 \indent Similarly, the summation tensor in eq. \ref{eq:QS_capacitance} is : \\
 
\begin{equation}
\boldsymbol{\Sigma}'' \equiv \begin{bmatrix}
\bovermat{ $N_{1}$ }{1 & 1 & \cdots} & \bovermat{$N_{2}$}{0 & 0 & \cdots} & \bovermat{$N_{b}$}{0 & 0 & \cdots} \\
0 & 0 & \cdots & 1 & 1 & \cdots & 0 & 0 & \cdots \\
0 & 0 & \cdots & 0 & 0 & \cdots & 1 & 1 & \cdots \\
\vdots & \vdots &  & \vdots & \vdots &  & \vdots & \vdots & \\
\mathbf{r}_{11} & \mathbf{r}_{12} & \cdots & 0 & 0 & \cdots & 0 & 0 & \cdots \\
0 & 0 & \cdots & \mathbf{r}_{21} & \mathbf{r}_{22} & \cdots & 0 & 0 & \cdots \\
0 & 0 & \cdots & 0 & 0 & \cdots & \mathbf{r}_{31} & \mathbf{r}_{32} & \cdots \\
\vdots & \vdots &  & \vdots & \vdots &  & \vdots & \vdots & \\
\end{bmatrix},
\end{equation}
\subsection*{Force and torques derivation}
\indent The potential energy of the system is expressed as a sum of products of rigid body moments and potential derivatives \cite{Jackson1999}
\begin{align}
U &= \frac{1}{2} \sum_\nu \left( Q_\nu \Psi_i - \mathbf{S}_i \cdot \mathbf{E}_0 \right) \\ \nonumber
&= \frac{1}{2} \mathbf{Q} \cdot \boldsymbol{\Psi}_0 + \frac{1}{2} \left[ \mathbf{q} \quad \boldsymbol{\Psi}-\boldsymbol{\Psi}_0 \right] \cdot \begin{bmatrix}
-\mathbf{r} \cdot \mathbf{E}_0 \\
\mathbf{Q}
\end{bmatrix}.
\end{align}
The induced charge distribution and rigid body potentials are computed from the inversion of equation \eqref{eq:saddle},
\begin{align} \label{eq:potenergy}
U &= \frac{1}{2} \mathbf{Q} \cdot \boldsymbol{\Psi}_0 \\ \nonumber
&+ \frac{1}{2} \left[ -\mathbf{r} \cdot \mathbf{E}_0 \quad \mathbf{Q} \right] \cdot \begin{bmatrix}
-\mathbf{M} & \boldsymbol{\Sigma}^T \\
\boldsymbol{\Sigma} & 0
\end{bmatrix}^{-1} \cdot \begin{bmatrix}
-\mathbf{r} \cdot \mathbf{E}_0 \\
\mathbf{Q}
\end{bmatrix}.
\end{align}
The force on an individual bead $\alpha$ is the negative derivative of the potential energy with respect to the bead position $\mathbf{x}_\alpha$, given all other beads remain fixed.  Because the rigid body charges and external potentials do not depend on the bead positions, the negative derivative of the first term in \eqref{eq:potenergy}, $- \nabla_{\mathbf{x}_\alpha}\mathbf{Q} \cdot \boldsymbol{\Psi}_0/2$, vanishes.  The product rule splits the negative derivative of the second term in \eqref{eq:potenergy} into three terms with the gradient acting only on one of the three matrices.  The first of these terms is,
\begin{align}
&-\Bigg( \nabla_{\mathbf{x}_\alpha} \left[ -\mathbf{r} \cdot \mathbf{E}_0 \quad \mathbf{Q} \right] \Bigg) \cdot \begin{bmatrix}
-\mathbf{M} & \boldsymbol{\Sigma}^T \\
\boldsymbol{\Sigma} & 0
\end{bmatrix}^{-1} \cdot \begin{bmatrix}
-\mathbf{r} \cdot \mathbf{E}_0 \\
\mathbf{Q}
\end{bmatrix} \\ \nonumber
&=  \left[ \nabla_{\mathbf{x}_\alpha} \mathbf{r} \cdot \mathbf{E}_0 \quad 0 \right] \cdot \begin{bmatrix}
\mathbf{q} \\
\boldsymbol{\Psi} - \boldsymbol{\Psi}_0
\end{bmatrix} \\ \nonumber
&= q_\alpha \mathbf{E}_0
\end{align}
and similarly for the term with the gradient applied to the third matrix.  Notice that only the $\alpha^\text{th }$ component of $\nabla_{\mathbf{x}_\alpha} \mathbf{r} \equiv [\nabla_{\mathbf{x}_\alpha} \mathbf{r}_1,\nabla_{\mathbf{x}_\alpha} \mathbf{r}_2,\dots,\nabla_{\mathbf{x}_\alpha} \mathbf{r}_\alpha,\dots] = [0,0,\dots,\mathbf{I},\dots]$ is nonzero, so only the $q_\alpha$ term survives the dot product.  The second term is,
\begin{widetext}
\begin{align}
 &-\left[ -\mathbf{r} \cdot \mathbf{E}_0 \quad \mathbf{Q} \right] \cdot \Bigg( \nabla_{\mathbf{x}_\alpha} \begin{bmatrix}
-\mathbf{M} & \boldsymbol{\Sigma}^T \\ 
\boldsymbol{\Sigma} & 0
\end{bmatrix}^{-1} \Bigg) \cdot \begin{bmatrix}
-\mathbf{r} \cdot \mathbf{E}_0 \\
\mathbf{Q}
\end{bmatrix} \\ \nonumber
& \qquad\qquad\qquad = \left[ -\mathbf{r} \cdot \mathbf{E}_0 \quad \mathbf{Q} \right] \cdot \begin{bmatrix}
-\mathbf{M} & \boldsymbol{\Sigma}^T \\
\boldsymbol{\Sigma} & 0
\end{bmatrix}^{-1} \cdot \begin{bmatrix}
-\nabla_{\mathbf{x}_\alpha}\mathbf{M} & \nabla_{\mathbf{x}_\alpha}\boldsymbol{\Sigma}^T \\
\nabla_{\mathbf{x}_\alpha} \boldsymbol{\Sigma} & 0
\end{bmatrix} \cdot \begin{bmatrix}
-\mathbf{M} & \boldsymbol{\Sigma}^T \\
\boldsymbol{\Sigma} & 0
\end{bmatrix}^{-1} \cdot \begin{bmatrix}
-\mathbf{r} \cdot \mathbf{E}_0 \\
\mathbf{Q}
\end{bmatrix} \\ \nonumber
& \qquad\qquad\qquad =\left[ \mathbf{q} \quad \boldsymbol{\Psi}-\boldsymbol{\Psi}_0 \right] \cdot \begin{bmatrix}
-\nabla_{\mathbf{x}_\alpha}\mathbf{M} & 0 \\
0 & 0
\end{bmatrix} \cdot \begin{bmatrix}
\mathbf{q} \\
\boldsymbol{\Psi} - \boldsymbol{\Psi}_0
\end{bmatrix} \\ \label{eq:productrule2}
& \qquad\qquad\qquad = - \mathbf{q} \cdot \nabla_{\mathbf{x}_\alpha} \mathbf{M} \cdot \mathbf{q} \nonumber
\end{align}
\end{widetext}
Notice that because the summation tensor is constant, its gradient is zero, $\nabla_{\mathbf{x}_\alpha} \boldsymbol{\Sigma} = 0$, and only the upper left block of the matrix survives.  Thus, the force on bead $\alpha$ is
\begin{equation} \label{eq:beadforce}
\mathbf{f}_\alpha = q_\alpha \mathbf{E}_0 - \frac{1}{2} \left( \nabla_{\mathbf{x}_\alpha} \mathbf{M} \right) : \mathbf{q}\mathbf{q} 
\end{equation}
Although the second term was written as $\mathbf{q} \cdot \nabla_{\mathbf{x}_\alpha} \mathbf{M} \cdot \mathbf{q}$ in \eqref{eq:productrule2}, rewriting the equations in index notation shows that both dot products should contract the vector indices of the charges with the indices of the potential tensor, leaving  the force with the same indices as the gradient.  Therefore the proper vector notation is the double dot product in \eqref{eq:beadforce}.  The force and torque on a rigid body can then be computed from the distribution of bead forces,
\begin{align}
\mathbf{F}_\nu &= \sum_\mu \mathbf{f}_{\nu\mu}, \\
\mathbf{L}_\nu &= \sum_\mu \mathbf{r}_{\nu\mu} \times \mathbf{f}_{\nu\mu}.
\end{align}   
These forces and torques serve as the input to a rigid body hydrodynamic integrator that computes the rigid body velocities and angular velocities of the composites by solving\cite{Swan2016, Wang2019A}
\begin{align}
\begin{bmatrix} \mathcal{M}^H & \Sigma'''^T \\
\Sigma''' & 0 \end{bmatrix} \cdot 
\begin{bmatrix} \mathbf{f}_c \\
\begin{pmatrix} \mathcal{U} \\
\boldsymbol{\Omega} \end{pmatrix} \end{bmatrix} =
\begin{bmatrix} 0 \\
\begin{pmatrix} \mathcal{F} \\
\mathcal{L} \end{pmatrix} \end{bmatrix}
\end{align}
where $\mathbf{f}_c$ is a list of $N$ constraint forces on each bead that hold the rigid composite together and $\Sigma'''$ is a summation tensor defined in reference \citenum{Swan2016}.  $\mathcal{U} = \left[ \mathbf{U}_1, \mathbf{U}_2, ... \mathbf{U}\right]^T$ is a list of rigid body velocities and similarly for the rigid body angular velocities $\boldsymbol{\Omega}$, rigid body forces $\mathcal{F}$, and rigid body torques $\mathcal{L}$.  Each bead $\nu\mu$ then moves with a velocity specified by the translational $\mathbf{U}_\nu$ and rotational $\boldsymbol{\Omega}_\nu$ velocities of the rigid body $\nu$ to which it belongs

\begin{equation}
\mathbf{u}_{\nu\mu} = \mathbf{U}_\nu + \boldsymbol{\Omega}_\nu \times \mathbf{r}_{\nu\mu}.
\end{equation}
\bibliography{aipsamp}

\end{document}